\documentclass[12pt,preprint]{aastex}

\shorttitle{PARTICLE ACCELERATION IN MAGNETIZED FILAMENTS}
\shortauthors{Honda \& Honda}

\begin{document}

\title{PARTICLE DIFFUSION AND ACCELERATION BY SHOCK WAVE IN
MAGNETIZED FILAMENTARY TURBULENCE}
\author{Mitsuru Honda}
\affil{Plasma Astrophysics Laboratory, Institute for Global Science,
Mie 519-5203, Japan}
\author{Yasuko S. Honda}
\affil{Kinki University Technical College, Mie 519-4395, Japan;
yasuko@ktc.ac.jp}

\begin{abstract}
We expand the off-resonant scattering theory for particle diffusion
in magnetized current filaments that can be typically compared to
astrophysical jets, including active galactic nucleus jets.
In a high plasma $\beta$ region where the directional bulk flow is
a free-energy source for establishing turbulent magnetic fields
via current filamentation instabilities, a novel version of quasi-linear
theory to describe the diffusion of test particles is proposed.
The theory relies on the proviso that the injected energetic particles are
not trapped in the small-scale structure of magnetic fields wrapping around
and permeating a filament but deflected by the filaments,
to open a new regime of the energy hierarchy mediated by a
transition compared to the particle injection.
The diffusion coefficient derived from a quasi-linear type
equation is applied to estimating the timescale for the stochastic
acceleration of particles by the shock wave propagating through the jet.
The generic scalings of the achievable highest energy of an
accelerated ion and electron, as well as of the characteristic
time for conceivable energy restrictions, are systematically presented.
We also discuss a feasible method of verifying
the theoretical predictions.
The strong, anisotropic turbulence reflecting cosmic filaments
might be the key to the problem of the acceleration mechanism of
the highest energy cosmic rays exceeding $100~{\rm EeV}$
($10^{20}~{\rm eV}$), detected in recent air shower experiments.
\end{abstract}


\section{INTRODUCTION}

Extremely high energy (EHE) cosmic rays beyond $100~{\rm EeV}$
have been observed in a couple of decades \citep{takeda98,abbasi04a},
but their origin still remains enigmatic.
In regard to the generation of the EHE particles, there are two
alternative schemes: the ``top-down'' scenario that hypothesizes
topological defects, Z-bursts, and so on, and the traditional
``bottom-up'' \citep[see, e.g.,][for a review]{olinto00}.
In the latter approach, we explore the candidate
celestial objects operating as a cosmic-ray ``Zevatron''
\citep{blandford00}, namely, an accelerator boosting
particle kinetic energy to ZeV ($10^{21}~{\rm eV}$) ranges.
By simply relating the celestial size to the gyroradius
for the typical magnetic field strength, one finds that the
candidates are restricted to only a few objects; these
include pulsars, active galactic nuclei (AGNs), radio galaxy
lobes, and clusters of galaxies \citep{hillas84,olinto00}.
In addition, gamma-ray bursters (GRBs) are known as
possible sources \citep{waxman95}.
As for the transport of EHE particles from the extragalactic sources,
within the GZK horizon \citep{greisen66,zatsepin66} the trajectory of
the particles (particularly protons) ought to suffer no significant
deflection due to the cosmological magnetic field, presuming
its strength of the order of $0.1~{\rm nG}$
\citep[see, e.g.,][for a review]{vallee04}.
According to a cross-correlation study \citep{farrar98}, some
super-GZK events seem to be well aligned with compact, radio-loud quasars.
Complementarily, self-correlation study is in progress,
showing small-scale anisotropy in the distribution of
the arrival direction of EHE primaries \citep{teshima03}.
More recently, the strong clustering has been confirmed,
as is consistent with the null hypothesis of isotropically
distributed arrival directions \citep{abbasi04b}.
At the moment, the interpretation of these results is
under active debate.

In the bottom-up scenario, the most promising mechanism for achieving EHE is
considered to be that of diffusive shock acceleration \citep[DSA;][]
{lagage83a,lagage83b,drury83}, which has been substantially studied
for solving the problems of particle acceleration in heliosphere
and supernova remnant (SNR) shocks
\citep[see, e.g.,][for a review]{blandford87}.
In general, it calls for the shock to be accompanied by
some kinds of turbulence that serve as the particle scatterers
\citep{krymskii77,bell78,blandford78}.
Concerning the theoretical modeling and its application to
extragalactic sources such as AGN jets, GRBs, and so forth,
it is still very important to know the actual magnetic field strength,
configuration, and turbulent state around the shock front.
At this juncture, modern polarization measurements by using very long
baseline interferometry began to reveal the detailed configuration of
magnetic fields in extragalactic jets, for example, the quite smooth
fields transverse to the jet axis \citep[1803+784:][]{gabuzda99}.
Another noticeable result is that within the current resolution,
a jet is envisaged as a bundle of {\it at least} a few filaments
(e.g., 3C\,84: \citealt{asada00}; 3C\,273: \citealt{lobanov01}),
as were previously confirmed in the radio arcs near the Galactic center
\citep[GC;][]{yusefzadeh84,yusefzadeh87}, as well as in the well-known
extragalactic jets
(e.g., Cyg\,A: \citealt{perley84}, \citealt{carilli96}; M87: \citealt{owen89}).

The morphology of filaments can be self-organized via the nonlinear
development of the electromagnetic current filamentation instability
\citep[CFI;][and references therein]{honda04} that breaks up a uniform beam
into many filaments, each carrying about net one unit current \citep{honda00}.
Fully kinetic simulations indicated that this subsequently led to the
coalescence of the filaments, self-generating significant toroidal
(transverse) components of magnetic fields \citep{honda00a,honda00b}.
As could be accommodated with this result, large-scale toroidal magnetic
fields have recently been discovered in the GC region \citep{novak03}.
Accordingly, we conjecture that a similar configuration appears in
extragalactic objects, particularly AGN jets \citep{hh04a}.
It is also pointed out that the toroidal fields could play a remarkable
role in collimating plasma flows \citep{honda02}.
Relating to this point, the AGN jets have narrow opening angles
of $\phi_{\rm oa}<10\degr$ in the long scales, although
in close proximity to the central engine the angles tend to spread
\citep[e.g., $\phi_{\rm oa}\approx 60\degr$ for the M87 jet;][]{junor99}.
Moreover, there is observational evidence that the internal pressures are
higher than the pressures in the external medium (e.g., 4C\,32.69:
\citealt{potash80}; Cyg\,A: \citealt{perley84}; M87: \citealt{owen89}).
These imply that the jets must be {\it self}-collimating and stably
propagating, as could be explained by the kinetic theory \citep{honda02}.

In the nonlinear stage of the CFI, the magnetized filaments
can often be regarded as strong turbulence that
more strongly deflects the charged particles.
When the shock propagation is allowed, hence, the particles
are expected to be quite efficiently accelerated for the
DSA scenario \citep{drury83,gaisser90}.
Indeed, such a favorable environment seems to be well established
in the AGN jets.
For example, in the filamentary M87 jet, some knots moving toward
the radio lobe exhibit the characteristics of shock discontinuity
\citep{biretta83,capetti97}, involving circumstantial evidence of
in situ electron acceleration \citep{meisenheimer96}.
As long as the shock accelerator operates for electrons, arbitrary
ions will be co-accelerated, providing that the ion abundance
in the jet is finite \citep[e.g.,][]{rawlings91,kotani96}.
It is, therefore, quite significant to study the feasibility
of EHE particle production in the filamentary jets with shocks:
this is just the original motivation for the current work.

This paper has been prepared to show a full derivation
of the diffusion coefficient for cosmic-ray particles
scattered by the magnetized filaments.
The present theory relies on a consensus that the kinetic energy
density (ram pressure) of the bulk plasma carrying currents is larger
than the energy density of the magnetic fields self-generated via
the CFI, likely comparable to the thermal pressure of the bulk.
That is, the flowing plasma as a reservoir of free energy
is considered to be in a high-$\beta$ state.
In a new regime in which the cosmic-ray particles interact off-resonantly
with the magnetic turbulence having no regular field, the quasi-linear
approximation of the kinetic transport equation is found to be
consistent with the condition that the accelerated particles
must be rather free from magnetic traps; namely, the
particles experience meandering motion.
It follows that the diffusion anisotropy becomes small.
Apparently, these are in contrast with the conventional quasi-linear
theory (QLT) for small-angle resonant scattering, according to which
one sets the resonance of the gyrating particles bound to a
mean magnetic field with the weak turbulence superimposed on
the mean field \citep{drury83,biermann87,longair92,hh04b}.
It is found that there is a wide parameter range in which
the resulting diffusion coefficient is smaller than that
from a simplistic QLT in the low-$\beta$ regime.
We compare a specified configuration of the filaments to an astrophysical
jet including AGN jets and discuss the correct treatment of what the
particle injection threshold in the present context could be.
We then apply the derived coefficient for calculations of the DSA
timescale and the achievable highest energy of accelerated particles
in that environment.
As a matter of convenience, we also show some generic scalings
of the highest energy, taking account of the conceivable energy
restrictions for both ions and electrons.

In order to systematically spell out the theoretical scenario, this paper is
divided into two major parts, consisting of the derivation of the diffusion
coefficient (\S~2) and its installation to the DSA model (\S~3).
We begin, in \S~2.1, with a discussion on the turbulent
excitation mechanism due to the CFI, so as to specify
a model configuration of the magnetized current filaments.
Then in \S~2.2, we explicitly formulate the equation that describes
particle transport in the random magnetic fluctuations.
In \S~2.3, the power-law spectral index of the magnetic
fluctuations is suggested for a specific case.
In \S~2.4, we write down the diffusion coefficients
derived from the transport equation.
In \S~3.1, we deal with the subject of particle injection, and
in \S~3.2, we estimate the DSA timescale, which is used to evaluate the
maximum energy of an accelerated ion (\S~3.3) and electron (\S~3.4).
Finally, \S~4 is devoted to a discussion of the feasibility and a summary.

\section[]{THEORY OF PARTICLE DIFFUSION IN MAGNETIC TURBULENCE
SUSTAINED BY ANISOTROPIC CURRENT FILAMENTS}

In what follows, given the spatial configuration of the magnetized
filaments of a bulk plasma jet, we derive the evolution equation
for the momentum distribution function of test particles, which is
linear to the turbulent spectral intensity, and then extract
an effective frequency for collisionless scattering and the
corresponding diffusion coefficient from the derived equation.

\subsection[]{\it Model Configuration of Magnetized Current Filaments}

Respecting the macroscopic transport of energetic particles in active galaxies,
there is direct/indirect observational evidence that they are
ejected from the central core of the galaxies and subsequently
transferred, through bipolar jets, to large-scale radio robes in which the
kinematic energy considerably dissipates \cite[e.g.,][]{biretta95,tashiro04}.
In this picture, it is expected that the directional plasma flows
will favorably induce huge currents in various aspects of the
above transport process (e.g., \citealt{appl92,conway93}; an
analogous situation also seems to appear in GRB jets:
e.g., \citealt{lyutikov03}).
Because of perfect conductivity in fully ionized plasmas, hot
currents driven in, e.g., the central engine prefer being quickly
compensated by plasma return currents.
This creates a pattern of the counterstreaming currents
that is unstable for the electromagnetic CFI.
As is well known, the pattern is also unstable to electrostatic
disturbances with the propagation vectors parallel to the
streaming direction, but in the present work, we eliminate
the longitudinal modes, so as to isolate the transverse CFI.
Providing a simple case in which the two uniform currents are
carried by electrons, the mechanism of magnetic field
amplification due to the CFI is explained as follows.
When the compensation of the counterpropagating electron currents
is disturbed in the transverse direction, magnetic repulsion
between the two currents reinforces the initial disturbance.
As a consequence, a larger and larger magnetic field is
produced as time increases.
For the Weibel instability as an example, the unstable mode is the
purely growing mode without oscillations \citep{honda04}, so the
temporal variation of magnetic fields is expected to be markedly
slow in the saturation regime (more on these is given
in \S\S~2.2 and 2.3).
Note that a similar pattern of quasi-static magnetic fields
can be also established during the collision of electron-positron
plasmas \citep{kazimura98,silva03} and in a shock front
propagating through an ambient plasma with/without initial
magnetic fields \citep{nishikawa03}.
These dynamics might be involved in the organization of the knotlike
features in the Fanaroff-Riley (FR) type~I radio jets, which appear to
be a shock caused by high-velocity material overtaking slower material
\citep[e.g.,][]{biretta83}.
Similarly, the cumulative impingement could also take place
around the hot spots of the FR type~II sources, which arguably
reflect the termination shocks.
In fact, the filamentary structure has been observed in
the hot spot region of a FR~II source \citep{perley84,carilli96}.

Furthermore, estimating the energy budget in many radio lobes
implies that the ram pressure of such current-carrying jets
is much larger than the energy density of the magnetic fields
\citep{tashiro04}; namely, the jet bulk can be regarded as a huge
reservoir of free energy.
In this regime, the ballistic motion is unlikely to be
affected by the self-generated magnetic fields,
as is actually seen in the linear feature of jets.
As a matter of fact, the GC region is known to arrange numerous
linear filaments, including nonthermal filaments \citep{yusefzadeh04}.

Taking these into consideration, we give a simple model
of the corresponding current--magnetic field system and
attempt to unambiguously distinguish the present system from the one
that appears in the low-$\beta$ plasmas hitherto well studied.
In Figure~1, for a given coordinate, we depict the configuration
of the linear current filaments and turbulent magnetic
fields of the bulk plasma.
Recalling that magnetic field perturbations develop in the direction
transverse to the initial currents \citep[e.g.,][]{honda04}, one
supposes the magnetic fields developed in the nonlinear phase to
be ${\bf B}=(B_{x},B_{y},0)$ \citep{montgomery79,medvedev99},
such that the vectors of zeroth-order current density
point in the directions parallel and antiparallel to the $z$-direction,
i.e., ${\bf J}\sim J{\hat{\bf z}}$, where the scalar $J$ ($\gtrless 0$)
is nonuniformly distributed on the transverse $x$-$y$ plane, while
uniformly distributed in the $z$-direction.
Note that for the fluctuating magnetic field vectors, we have used
the simple character (${\bf B}$) without any additional symbol such as
``$\delta$,'' since the establishment of no significant regular component
is expected, and simultaneously, ${\bf J}$ ($\sim\nabla\times{\bf B}$)
well embodies the quasi-static current filaments in the zeroth order.
For convenience, hereafter, the notations ``parallel''
($\parallel$) and ``transverse'' ($\perp$) are referred to
as the directions with respect to the linear current filaments
aligned in the $z$-axis, as they are well defined reasonably
(n.b. in the review of \S~2.4, $\parallel_{b}$ and $\perp_{b}$
with the subscript ``$b$'' refer to a mean magnetic field line).
It is mentioned that the greatly fluctuating transverse fields could be
reproduced by some numerical simulations \citep[e.g.,][]{lee73,nishikawa03}.
In an actual filamentary jet, a significant reduction of polarization
has been found in the center, which could be ascribed to the cancellation
of the small-scale structure of magnetic fields \citep{capetti97},
compatible with the present model configuration.
In addition, there is strong evidence that random fields accompany GRB jets
\citep[e.g.,][]{greiner03}.
The arguments expanded below highlight the transport properties
of test particles in such a bulk environment, that is,
in a forest of magnetized current filaments.

\subsection[]{\it The Quasi-linear Type Equation for Cosmic-Ray Transport}

We are particularly concerned with the stochastic diffusion of the
energetic test particles injected into the magnetized current filaments
(for the injection problem, see the discussion in \S~3.1).
As a rule, the Vlasov equation is appropriate for describing the
collisionless transport of the relativistic particles in the turbulent
magnetic field, ${\bf B}({\bf r},t)$, where ${\bf r}=(x,y)$ and
the slow temporal variation has been taken into consideration.
Transverse electrostatic fields are ignored, since
they preferentially attenuate over long timescales,
e.g., in the propagation time of jets (see \S~2.3).
The temporal evolution of the momentum distribution function
for the test particles, $f_{\bf p}$, can then be described as

\begin{equation}
{{\rm D}f_{\bf p}\over{{\rm D}t}}={\partial\over{\partial t}}f_{\bf p}
+\left({\bf v}\cdot{\partial\over{\partial{\bf r}}}\right)f_{\bf p}
+{q\over c}\left[\left({\bf v}\times{\bf B}\right)
\cdot{\partial\over{\partial{\bf p}}}\right]f_{\bf p}=0
\label{eqn:1}
\end{equation}

\noindent
for arbitrary particles. Here $q$ is the particle charge,\footnote {For
example, $q=-|e|$ for electrons, $q=|e|$ for positrons,
and $q=Z|e|$ for ions or nuclei, where $e$ and $Z$ are the
elementary charge and the charge number, respectively.}
$c$ is the speed of light, and the other notations are standard.
We decompose the total distribution function
into the averaged and fluctuating part,
$f_{\bf p}=\left<f_{\bf p}\right>+\delta f_{\bf p}$, and consider
the specific case in which from a macroscopic point of view, the vector
${\bf B}=(B_{x},B_{y},0)$ is randomly distributed on the
transverse $x$-$y$ plane \citep{montgomery79,medvedev99}.
Taking the ensemble average of equation~(\ref{eqn:1}),
$\left<{\rm D}f_{\bf p}/{{\rm D} t}\right>=0$, then yields

\begin{equation}
{\partial\over{\partial t}}\left<f_{\bf p}\right>
+\left({\bf v}\cdot{\partial\over{\partial{\bf r}}}\right)
\left<f_{\bf p}\right>=-{q\over c}
\left<\left[\left({\bf v}\times{\bf B}\right)
\cdot{\partial\over{\partial{\bf p}}}\right]\delta f_{\bf p}\right>,
\label{eqn:2}
\end{equation}

\noindent
where we have used $\left<{\bf B}\right>\simeq {\bf 0}$.
Taking account of no mean field implies that we do not
invoke the gyration and guiding center motion of the particles.
Subtracting equation~(\ref{eqn:2}) from equation (\ref{eqn:1})
and picking up the term linear in fluctuations, viz., employing
the conventional quasi-linear approximation, we obtain

\begin{equation}
{\partial\over{\partial t}}\delta f_{\bf p}
+\left({\bf v}\cdot{\partial\over{\partial{\bf r}}}\right)\delta f_{\bf p}=
-{q\over c}\left[\left({\bf v}\times{\bf B}\right)
\cdot{\partial\over{\partial{\bf p}}}\right]\left<f_{\bf p}\right>.
\label{eqn:3}
\end{equation}

\noindent
As usual, equation~(\ref{eqn:3}) is valid for
$\left<f_{\bf p}\right>\gg|\delta f_{\bf p}|$ \citep{landau81}.
As shown in \S~3.1, this condition turns out to be consistent
with the aforementioned implication that the injected test particles
must be free from the small-scale magnetic traps embedded in the bulk.
Relating to this, note that to remove the ambiguity of terminologies,
the injected, energetic test particles obeying $f_{\bf p}$
are just compared to the cosmic rays that are shown below to be
diffusively accelerated owing to the present scenario.
Within the framework of the test particle approximation,
the back-reaction of the slow spatiotemporal change of
$\left<f_{\bf p}\right>$ to the modulation of ${\bf B}$
(sustained by the bulk) is ignored, in contrast to the case for
SNR environments, where such effects often become nonnegligible
\citep[e.g.,][]{bell04}.

In general, the vector potential conforms to ${\bf B}=\nabla\times{\bf A}$
and $\nabla\cdot{\bf A}=0$.
For the standard, plane wave approximation, we carry out
the Fourier transformation of the fluctuating components
for time and the transverse plane:

\begin{equation}
\delta f_{\bf p}({\bf r},t)=\int\delta f_{{\bf p},{\cal K}}
e^{i\left[\left({\bf k}\cdot{\bf r}\right)-\omega t\right]}{\rm d}^3{\cal K},
\label{eqn:4}
\end{equation}
\begin{equation}
{\bf A}({\bf r},t)=\int{\bf A}_{\cal K}
e^{i\left[\left({\bf k}\cdot{\bf r}\right)-\omega t\right]}{\rm d}^3{\cal K},
\label{eqn:5}
\end{equation}
\begin{equation}
{\bf B}({\bf r},t)=\int{\bf B}_{\cal K}
e^{i\left[\left({\bf k}\cdot{\bf r}\right)-\omega t\right]}{\rm d}^3{\cal K},
\label{eqn:6}
\end{equation}

\noindent
and ${\bf B}_{\cal K}=i{\bf k}\times{\bf A}_{\cal K}$,
where $i=\sqrt{-1}$, ${\cal K}=\left\{{\bf k},\omega\right\}$,
and ${\rm d}^3{\cal K}={\rm d}^2{\bf k}{\rm d\omega}$.
The given magnetic field configuration follows
${\bf A}=A{\hat{\bf z}}$ (${\bf A}_{\cal K}=A_{\cal K}{\hat{\bf z}}$)
and ${\bf k}\perp{\hat{\bf z}}$.
As illustrated in Figure~1, the scalar quantity $A$ ($\gtrless 0$)
is also random on the transverse plane, with no mean value.
Making use of equations~(\ref{eqn:4})--(\ref{eqn:6}),
equation~(\ref{eqn:3}) can be transformed into

\begin{equation}
\delta f_{{\bf p},{\cal K}}={q\over c}
\left[\omega -\left({\bf k}\cdot{\bf v}\right)\right]^{-1}
\left[{\bf v}\times\left({\bf k}\times{\bf A}_{\cal K}\right)\right]
\cdot{{\partial\left< f_{\bf p}\right>}\over{\partial{\bf p}}}.
\label{eqn:7}
\end{equation}

\noindent
On the other hand, the right-hand side (RHS) of
equation~(\ref{eqn:2}) can be written as

\begin{equation}
{\rm RHS}=-i{q\over c}\left<\int{\rm d}^3{\cal K}^{\prime}
e^{i\left[\left({\bf k}^{\prime}\cdot{\bf r}\right)-\omega^{\prime} t\right]}
\left\{\left[{\bf v}\times\left({\bf k}^{\prime}\times
{\bf A}_{{\cal K}^{\prime}}\right)\right]
\cdot{\partial\over{\partial{\bf p}}}\right\}\delta f_{\bf p}\right>.
\label{eqn:8}
\end{equation}

\noindent
Substituting equation~(\ref{eqn:4}) (involving eq.~[\ref{eqn:7}]) into
equation~(\ref{eqn:8}), equation~(\ref{eqn:2}) can be expressed as

\begin{eqnarray}
{{{\rm d}\left<f_{\bf p}\right>}\over{{\rm d}t}}=-i{q^2\over c^2}
\left<\int{\rm d}^3{\cal K}{\rm d}^3{\cal K}^{\prime}
e^{i\left\{\left[\left({\bf k}+{\bf k}^{\prime}\right)\cdot{\bf r}\right]
-\left(\omega+\omega^{\prime}\right)t\right\}}\right. \nonumber \\
\left. \left[{\bf v}\times\left({\bf k}^{\prime}\times
{\bf A}_{{\cal K}^{\prime}}\right)\right]
\cdot{\partial\over{\partial{\bf p}}}
\left\{{{\left[{\bf v}\times\left({\bf k}\times
{\bf A}_{\cal K}\right)\right]}\over
{\omega-\left({\bf k}\cdot{\bf v}\right)}}
\cdot{\partial\over{\partial{\bf p}}}
\right\}\left< f_{\bf p}\right>\right>,
\label{eqn:9}
\end{eqnarray}

\noindent
where the definition of the total derivative,
${\rm d}/{\rm d}t\equiv\partial/{\partial t}+{\bf v}\cdot
\left({\partial/{\partial{\bf r}}}\right)$, has been introduced.
As for the integrand of equation~(\ref{eqn:9}),
it may be instructive to write down the vector identity of

\begin{equation}
{\bf v}\times\left({\bf k}^{(\prime)}\times{\bf A}_{{\cal K}^{(\prime)}}\right)
=\left({\bf v}\cdot{\bf A}_{{\cal K}^{(\prime)}}\right){\bf k}^{(\prime)}
-\left({\bf k}^{(\prime)}\cdot{\bf v}\right){\bf A}_{{\cal K}^{(\prime)}}.
\label{eqn:10}
\end{equation}

\noindent
From the general expression of equation~(\ref{eqn:9}),
we derive an effective collision frequency that stems from
fluctuating field-particle interaction, as shown below.

For convenience, we decompose the collision integral
(RHS of eq.~[\ref{eqn:9}]) including the scalar products,
$\cdot\left(\partial/\partial{\bf p}\right)$, into the four parts:

\begin{equation}
{{{\rm d}\left<f_{\bf p}\right>}\over{{\rm d}t}}=\sum_{i,j}I_{ij},
\label{eqn:11}
\end{equation}

\noindent
where $i,j=1,2$ and

\begin{equation}
I_{ij}
\equiv -i{q^2\over c^2}\left<\int{\rm d}^3{\cal K}{\rm d}^3{\cal K}^{\prime}
\cdots{\partial\over{\partial p_{i}}}
\cdots{\partial\over{\partial p_{j}}}\left< f_{\bf p}\right>\right>.
\label{eqn:12}
\end{equation}

\noindent
In the following notations, the subscripts ``1'' and ``2''
indicate the parallel ($\parallel$) and perpendicular ($\perp$)
direction to the current filaments, respectively.
Below, as an example, we investigate the contribution from the integral
$I_{11}$ (see Appendix for calculation of the other components).
For the purely parallel diffusion involving the partial
derivative of only $\partial/\partial p_{\parallel}$,
the first term of the RHS of equation~(\ref{eqn:10})
does not make a contribution to equation~(\ref{eqn:9}).

In the ordinary case in which the random fluctuations are
stationary and homogeneous, the correlation function
has its sharp peak at $\omega=-\omega^{\prime}$ and
${\bf k}=-{\bf k}^{\prime}$ \citep{tsytovich95}, that is,

\begin{equation}
\left< A_{\cal K}A_{{\cal K}^{\prime}}\right> =|A|_{{\bf k},\omega}^{2}
\delta({\bf k}+{\bf k}^{\prime})\delta(\omega+\omega^{\prime}),
\label{eqn:13}
\end{equation}

\noindent
where the Dirac $\delta$-function has been used.
Here note the relation of
$|A|_{{\bf k},\omega}^{2}=|A|_{{\bf -k},-\omega}^{2}$ because we have
$A_{{\bf -k},-\omega}=A_{{\bf k},\omega}^{*}$, where the superscript
asterisk indicates the complex conjugate; this is valid as far as
${\bf A}({\bf r},t)$ is real, i.e., ${\bf B}({\bf r},t)$ is observable.
By using equation~(\ref{eqn:13}), the integral component
$I_{11}$ can be expressed as

\begin{equation}
I_{11}=i{q^{2}\over c^{2}}
\int{\rm d}^{2}{\bf k}{\rm d}\omega|A|_{{\bf k},\omega}^{2}
\left({\bf k}\cdot{\bf v}\right)
{\partial\over{\partial p_{\parallel}}}
\left[{{{\bf k}\cdot{\bf v}}
\over{\omega-\left({\bf k}\cdot{\bf v}\right)}}
{\partial\over{\partial p_{\parallel}}}\left<f_{\bf p}\right>\right],
\label{eqn:14}
\end{equation}

\noindent
where the relation of
${\bf A}_{\cal K}\cdot\left(\partial/\partial{\bf p}\right)
=A_{\cal K}\left(\partial/\partial p_{\parallel}\right)$
has been used.
In order to handle the resonant denominator of equation~(\ref{eqn:14}),
we introduce the causality principle of
$\lim_{\epsilon\rightarrow +0}\left[\omega-
\left({\bf k}\cdot{\bf v}\right)+i\epsilon\right]^{-1}
\rightarrow{\cal P}\left[\omega-\left({\bf k}\cdot{\bf v}\right)\right]^{-1}
-i\pi\delta\left[\omega-\left({\bf k}\cdot{\bf v}\right)\right]$,
where ${\cal P}$ indicates the principal value \citep{landau81}.
One can readily confirm that the real part of the resonant
denominator does not contribute to the integration.
Thus, we have

\begin{equation}
I_{11}={{\pi q^{2}}\over c^{2}}
\int{\rm d}^{2}{\bf k}{\rm d}\omega|A|_{{\bf k},\omega}^{2}
\left({\bf k}\cdot{\bf v}\right)
{\partial\over{\partial p_{\parallel}}}
\left\{\delta\left[\omega-\left({\bf k}\cdot{\bf v}\right)\right]
\left({\bf k}\cdot{\bf v}\right)
{\partial\over{\partial p_{\parallel}}}
\left<f_{\bf p}\right>\right\}.
\label{eqn:15}
\end{equation}

\noindent
Equation~(\ref{eqn:15}) shows the generalized form of the
quasi-linear equation, allowing $|A|_{{\bf k},\omega}^{2}$
to be arbitrary functions of ${\bf k}$ and $\omega$.\footnote{For the
case in which the unstable mode is a wave mode with
$\omega_{\bf k}\neq0$, the frequency dependence of the correlation function
can be summarized in the form of $|{\cal F}|_{{\bf k},\omega}^{2}=
|{\cal F}|_{\bf k}^{2}\delta\left(\omega-\omega_{\bf k}\right)+
|{\cal F}|_{\bf -k}^{2}\delta\left(\omega+\omega_{\bf k}\right)$,
which is valid for weak turbulence
concomitant with a scalar or vector potential ${\cal F}$.
However, this is not the case considered here.
The free-energy source that drives instability is now
current flows; thereby, unstable modes without oscillation
(or with quite slow oscillation) can be excited.}
In the present circumstances, a typical unstable mode of the
CFI is the purely growing Weibel mode with $\omega=0$ in
collisionless regimes, although in a dissipative regime
the dephasing modes with a finite but small value of
$\omega=\pm\Delta\omega_{\bf k}$ are possibly excited \citep{honda04}.
In the latter case, the spectral lines will be broadened
in the nonlinear phase.
Nevertheless, it is assumed that the spectrum still retains
the peaks around $\omega\approx\pm\Delta\omega$,
accompanied by their small broadening of the same order,
where $|\Delta\omega|\ll\gamma_{\bf k}\sim\omega_{\rm p}$,
and $\gamma_{\bf k}$ and $\omega_{\rm p}/(2\pi)$ are the
growth rate and the plasma frequency, respectively.
In the special case reflecting the purely growing mode,
the spectrum retains a narrow peak at $\omega =0$ with
$|\Delta\omega|\sim 0$ \citep{montgomery79}.
Apparently, the assumed quasi-static properties are in
accordance with the results of the fully kinetic simulations
\citep{kazimura98,honda00a}, except for a peculiar temporal
property of the rapid coalescence of filaments.
Accordingly, here we employ an approximate expression of

\begin{equation}
|A|_{{\bf k},\omega}^{2}\sim
|A|_{\bf k}^{2}\delta\left(\omega-\Delta\omega\right)+
|A|_{\bf -k}^{2}\delta\left(\omega+\Delta\omega\right),
\label{eqn:16}
\end{equation}

\noindent
where $|A|_{\bf k}^{2}=|A|_{\bf -k}^{2}$.
Note that when taking the limit of $|\Delta\omega|\rightarrow 0$,
equation~(\ref{eqn:16}) degenerates into
$|A|_{{\bf k},\omega}^{2}\sim 2|A|_{\bf k}^{2}\delta\left(\omega\right)$.

Substituting equation~(\ref{eqn:16}) into equation~(\ref{eqn:15}) yields

\begin{equation}
I_{11}\sim{{2\pi q^{2}}\over c^{2}}\int{\rm d}^{2}{\bf k}|A|_{\bf k}^{2}
\left({\bf k}\cdot{\bf v}\right)
{\partial\over{\partial p_{\parallel}}}
\left\{\delta\left[\Delta\omega -\left({\bf k}\cdot{\bf v}\right)\right]
\left({\bf k}\cdot{\bf v}\right)
{\partial\over{\partial p_{\parallel}}}
\left<f_{\bf p}\right>\right\}.
\label{eqn:17}
\end{equation}

\noindent
Furthermore, we postulate that the turbulence is isotropic
on the transverse plane, though still, of course, allowing
anisotropy of the vectors ${\bf A}$ parallel to the $z$-axis.
Equation~(\ref{eqn:17}) can be then cast to

\begin{equation}
I_{11}\sim{{2\pi q^{2}}\over c^{2}}
v_{\perp}{\partial\over{\partial p_{\parallel}}}
\int {{\rm d}\theta\over{2\pi}}\cos^{2}\theta
\int{\rm d}k 2\pi k
\delta\left(\Delta\omega -kv_{\perp}\cos\theta\right)
{k^{2}|A|_{k}^{2}}v_{\perp}{\partial\over{\partial p_{\parallel}}}
\left<f_{\bf p}\right>,
\label{eqn:18}
\end{equation}

\noindent
where $k=|{\bf k}|$, $v_{\perp}=|{\bf v}_{\perp}|$, and
${\bf k}\cdot{\bf v}={\bf k}\cdot\left({\bf v}_{\perp}+v_{\parallel}
{\hat{\bf z}}\right)=kv_{\perp}\cos\theta$.

As concerns the integration for $\theta$, we see that the
contribution from the marginal region of the smaller $|\cos\theta|$,
reflecting narrower pitch angle, is negligible.
In astrophysical jets, the pitch angle distribution for
energetic particles still remains unresolved, although
the distribution itself is presumably unimportant.
Hence, at the moment it may be adequate to simply take an angular average,
considering, for heuristic purposes, the contribution from the range of
$|\cos\theta|\sim O(1)\gg\epsilon$ for a small value of $\epsilon$.
If one can choose
$\epsilon\gtrsim|\Delta\omega|/(k_{\rm min}v_{\perp})$, the above relation,
$\epsilon\ll |\cos\theta|$, reflects the off-resonant interaction,
i.e., $|\Delta\omega|\ll |{\bf k}\cdot{\bf v}|$.
The minimum wavenumber, $k_{\rm min}$, is typically of the order
of the reciprocal of the finite system size, which is, in the present
circumstances, larger than the skin depth $c/\omega_{\rm p}$.
These ensure the aforementioned relation of
$|\Delta\omega|\ll \omega_{\rm p}$ (or $|\Delta\omega|\sim 0$).
In addition, the off-resonance condition provides an approximate
expression of $\delta\left(\Delta\omega -kv_{\perp}\cos\theta\right)
\sim\left(kv_{\perp}|\cos\theta|\right)^{-1}$.
Using this expression, the integral for the angular average
can be approximated by
$\sim\int_{0}^{2\pi}\cos^{2}\theta/(2\pi|\cos\theta|){\rm d}\theta=2/\pi$.
This is feasible, on account of the negligible contribution from the angle of
$|\cos\theta|\lesssim\epsilon$.
Then equation~(\ref{eqn:18}) reduces to

\begin{equation}
I_{11}\sim{{16\pi q^{2}}\over c^{2}}v_{\perp}
{\partial^2 \over{\partial p_{\parallel}^{2}}}\left<f_{\bf p}\right>
\int_{k_{\rm min}}^{k_{\rm max}}{{{\rm d}k}\over k}I_{k},
\label{eqn:19}
\end{equation}

\noindent
where we have defined the modal energy density
(spectral intensity) of the quasi-static turbulence by
$I_{k}\equiv2\pi k\left(k^{2}|A|_{k}^{2}/4\pi\right)$, such that
the magnetic energy density in the plasma medium can be evaluated by
$u_{\rm m}\simeq\left<|{\bf B}|^{2}\right>/8\pi
=\int_{k_{\rm min}}^{k_{\rm max}}I_{k}{\rm d}k$.

\subsection[]{\it Spectral Intensity of the Transverse Magnetic Fields}

The energy density of the quasi-static magnetic fields, $u_{\rm m}$,
likely becomes comparable to the thermal pressure of the filaments
\citep{honda00a,honda00b,honda02}.
When exhibiting such a higher $u_{\rm m}$ level, the bulk plasma
state may be regarded as the strong turbulence; but recall that
in the nonlinear CFI, the frequency spectrum with a sharp peak
at $\omega =0$ is scarcely smoothed out, since significant 
mode-mode energy exchanges are unexpected.
This feature is in contrast to the ordinary magnetohydrodynamic (MHD)
and electrostatic turbulence, in which a larger energy density
of fluctuating fields would involve modal energy transfer.
One of the most remarkable points is that as long as
the validity condition of the quasi-linear approximation,
$\left<f_{\bf p}\right>\gg |\delta f_{\bf p}|$, is satisfied
(for details, see \S~3.1), the present off-resonant scattering
theory covers even the strong turbulence regime.
That is, the theory, which might be classified into an extended
version of the QLT, does not explicitly restrict the magnetic
turbulence to be weak (for instruction, Tsytovich \& ter~Haar [1995]
have considered a generalization of the quasi-linear equation in
regard to its application to strong electrostatic turbulence).
Apparently, this is also in contrast to the conventional
QLT for small-angle resonant scattering, which invokes
a mean magnetic field (well defined only for the case
in which the turbulence is weak) in ordinary low-$\beta$ plasmas.

In any case, in equation~(\ref{eqn:19}) we specify the spectral
intensity of the random magnetic fields, which are established
via the aforementioned mechanism of the electromagnetic CFI.
The closely related analysis in the nonlinear regime was first performed by
\citet{montgomery79}, for a simple case in which two counterstreaming
electron currents compensate for a uniform, immobile ion background.
In the static limit of $\omega\rightarrow 0$, they have derived
the modal energy densities of fluctuating electrostatic and
magnetic fields, by using statistical mechanical techniques.
They predicted the accumulation of magnetic energy at long wavelengths,
consistent with the corresponding numerical simulation \citep{lee73}.
It was also shown that at long wavelengths, the energy density of a
transverse electrostatic field was comparable to the thermal energy density.
However, when allowing ion motions, such an electrostatic field is
found to attenuate significantly, resulting in equipartition of the energy
into magnetic and thermal components \citep{honda00a,honda00b}.
That is why we have neglected the electrostatic field in
equation~(\ref{eqn:1}).

When the spectral intensity of the magnetic fluctuations
can be represented by a power-law distribution of the form

\begin{equation}
I_{k}\propto k^{-\alpha},
\label{eqn:20}
\end{equation}

\noindent
we refer to $\alpha$ as spectral index.
\citet{montgomery79} found that for the transverse magnetic
fields accompanying anisotropic current filaments,
the spectral index could be approximated by
$\alpha\approx 2$ in a wide range of $k$, that is,

\begin{equation}
I_{k}\propto k^{-2}.
\label{eqn:21}
\end{equation}

\noindent
Note that the spectral index is somewhat larger than
$\alpha_{\rm MHD}\simeq 1-5/3$ for the classical
MHD context \citep{kolmogorov41,bohm49,kraichnan65}.
The larger index is rather consistent with the observed trends of
softening of filamentary turbulent spectra in extragalactic jets
\citep[e.g., $\alpha\simeq 2.6$ in Cyg~A;][and references therein]{carilli96}.
Although the turbulent dissipation actually involves the truncation
of $I_{k}$ in the short-wavelength regions, we simply take
$k_{\rm max}\rightarrow\infty$, excluding the complication.
Using equation~(\ref{eqn:20}) and the expression of the
magnetic energy density $u_{\rm m}$, we find the relation of

\begin{equation}
\int_{k_{\rm min}}^{\infty}{{{\rm d}k}\over k}I_{k}=
{1\over k_{\rm min}}{{\alpha-1}\over \alpha}
{{\left<|{\bf B}|^{2}\right>}\over{8\pi}}
\label{eqn:22}
\end{equation}

\noindent
for $\alpha >1$.
The spectral details for individual jets \citep[such as the
bend-over scales of $I_{k}$, correlation length, and so on; e.g.,
for the heliosphere, see][]{zank98,zank04} will render the
integration of equation~(\ref{eqn:22}) more precise, but the
related observational information has been poorly updated thus far.
For the present purpose, we simply use equation~(\ref{eqn:22}),
setting $k_{\rm min}=\pi/R$, where $R$ stands for the radius of
the jet, which is actually associated with the radius of a bundle of
filaments of various smaller radial sizes \citep[e.g.,][]{owen89}.
This ensures that the coherence length of the fluctuating force,
$\sim k^{-1}$, is small compared with a characteristic system size,
i.e., the transverse size, as is analogous to the restriction
for use of the conventional QLT.

\subsection[]{\it The Diffusion Coefficients}

In order to evaluate the diffusion coefficients of test particles,
one needs to specify the momentum distribution function,
$\left<f_{\bf p}\right>$, in equation~(\ref{eqn:19}).
As is theoretically known, the Fermi acceleration mechanisms lead to
the differential spectrum of ${\rm d}n/{\rm d}E\propto E^{-\beta}$
[or $n(>E)\propto E^{-\beta+1}$; \citealt{gaisser90}],
where ${\rm d}n(E)$ defines the density of particles
with kinetic energy between $E$ and $E+{\rm d}E$.
For the first-order Fermi mechanism involving nonrelativistic
shock with its compression ratio of $r\leq 4$, the power-law index
reads $\beta=\left( r+2\right) /\left( r-1\right)\geq 2$,
accommodated by the observational results.
With reference to these, we have the momentum distribution function
of $\left<f_{\bf p}\right>\propto |{\bf p}|^{-(\beta +2)}$
for the ultrarelativistic particles having $E=|{\bf p}|c$,
such that in the isotropic case, the differential quantity
$\left<f_{\bf p}\right>|{\bf p}|^{2}{\rm d}|{\bf p}|/(2\pi^{2})$
corresponds to ${\rm d}n(E)$ defined above \citep[e.g.,][]{blandford78}.
Then, in equation~(\ref{eqn:19}) the partial derivative of
the distribution function can be estimated as
$\partial^{2}\left< f_{\bf p}\right>/\partial p_{\parallel}^{2}\sim
(\beta +2)[\left(\beta +3\right)(p_{\parallel}/|{\bf p}|)^{2}
-(p_{\perp}/|{\bf p}|)^{2}](c^{2}/E^{2})\left< f_{\bf p}\right>$,
where we have used $|{\bf p}|^{2}=p_{\parallel}^{2}+p_{\perp}^{2}$.
Making use of this expression and equation~(\ref{eqn:22}),
equation~(\ref{eqn:19}) can be arranged in the form of
$I_{11}\sim\nu_{11}\left< f_{\bf p}\right>$.
Here $\nu_{11}$ reflects an effective collision frequency
related to the purely parallel diffusion in momentum space, to give

\begin{equation}
\nu_{11}={{2\left(\alpha -1\right)\left(\beta +2\right)
\left[\left(\beta +3\right)
\psi_{1}^{2}-\psi_{2}^{2}\right]\psi_{2}}\over
{\pi\alpha}}{{cq^{2}B^{2}R}\over{E^{2}}},
\label{eqn:23}
\end{equation}

\noindent
where we have used the definitions of
$B^{2}\equiv\left<|{\bf B}|^{2}\right>$, and
$\psi_{1}\equiv p_{\parallel}/|{\bf p}|\gtrless 0$ and
$\psi_{2}\equiv p_{\perp}/|{\bf p}|>0$, whereby $\sum_{i}\psi_{i}^{2}=1$.

Similarly, one can calculate the other components of
the integral $I_{ij}$ as outlined in the Appendix
and arrange them in the form of
$I_{ij}\sim\nu_{ij}\left< f_{\bf p}\right>$.
As a result, we obtained

\begin{equation}
\nu_{22}=
{{2\left(\alpha -1\right)\left(\beta +2\right)\left(\beta +4\right)
\psi_{1}^{2}\psi_{2}}\over{\pi\alpha}}{{cq^{2}B^{2}R}\over{E^{2}}},
\label{eqn:24}
\end{equation}

\noindent
and
\begin{eqnarray}
\nu_{12}&=&-\nu_{11}, \nonumber \\
\nu_{21}&=&-\nu_{22}.
\label{eqn:25}
\end{eqnarray}

\noindent
As would be expected, we confirm a trivial relation of
${{\rm d}\left<f_{\bf p}\right>}/{{\rm d}t}=\sum_{i,j}I_{ij}
\sim\sum_{i,j}\nu_{ij}\left<f_{\bf p}\right>=0$,
stemming from the orthogonality in the RHS of equation~(\ref{eqn:2}).

Now we estimate the spatial diffusion coefficients in an ad hoc
manner: $\kappa_{ij}\sim c^{2}\psi_{i}\psi_{j}/(2\nu_{ij})$.
It is then found that the off-diagonal components,
$\kappa_{12}$ and $\kappa_{21}$, include the factor of
${\rm sgn}(\psi_{1}\gtrless 0)=\pm 1$, implying that
these components vanish for an average.
For $\psi_{1}^{2}={1\over 3}$ and $\psi_{2}^{2}={2\over 3}$
reflecting the momentum isotropy, the diffusion coefficients
can be summarized in the following tensor form:

\begin{eqnarray}
\left(
\begin{array}{cc}
        \kappa_{\parallel} & 0 \\
        0 & \kappa_{\perp} \\
\end{array}
\right)
&\sim&{\sqrt{6}\pi\alpha\over{8\left(\alpha -1\right)}}
{{cE^{2}}\over{q^{2}B^{2}R}}\left[
\begin{array}{cc}
        {1\over{\left(\beta +1\right)\left(\beta +2\right)}} &
        0 \\
        0 &
        {2\over {\left(\beta +2\right)\left(\beta +4\right)}} \\
\end{array}
\right],
\label{eqn:26}
\end{eqnarray}

\noindent
where $\kappa_{\parallel}\equiv\kappa_{11}$ and
$\kappa_{\perp}\equiv\kappa_{22}$.
The perpendicular component can be expressed as
$\kappa_{\perp}={\tilde\kappa}\kappa_{\parallel}$, where
${\tilde\kappa}\equiv 2(\beta +1)/(\beta +4)$.
Note the allowable range of $1\leq{\tilde\kappa}<2$
for $\beta\geq 2$; particularly, ${\tilde\kappa}\approx 1$
for the expected range of $\beta\approx 2-3$.

It may be instructive to compare the diffusion coefficient
of equation~(\ref{eqn:26}) with that derived from the
previously suggested theories including the QLT.
In weakly turbulent low-$\beta$ plasmas, the mean magnetic field
with its strength ${\bar B}$, which can bind charged particles
and assign the gyroradius of $r_{\rm g}=E/(|q|{\bar B})$,
provides a well-defined direction along the field line; therefore,
in the following discussion, we refer, for convenience, to
${\parallel}_{b}$ and ${\perp}_{b}$ as the parallel and perpendicular
directions to the mean magnetic field, respectively.
For a simplistic QLT, one sets an ideal environment in which the
turbulent Alfv\'en waves propagating along the mean field
line resonantly scatter the bound particles, when
$k_{{\parallel}_{b}}^{-1}\sim r_{\rm g}$, where $k_{{\parallel}_{b}}$
is the parallel wavenumber \citep{drury83,longair92}.
Assuming that the particles interact with the waves in the
inertial range of the turbulent spectrum with its index $\alpha_{b}$,
the parallel diffusion coefficient could be estimated as
\citep{biermann87,muecke01}

\begin{equation}
\kappa_{{\parallel}_{b}}\sim
{1\over{3\left(\alpha_{b}-1\right)\eta_{b}}}
{{cr_{\rm g}}\over
{\left(k_{{\parallel}_{b},{\rm min}}r_{\rm g}\right)^{\alpha_{b}-1}}}
\label{eqn:27}
\end{equation}

\noindent
for $\alpha_{b}\neq 1$ and
$r_{\rm g}\leq k_{{\parallel}_{b},{\rm min}}^{-1}$,
where $k_{{\parallel}_{b},{\rm min}}^{-1}$ reflects the
correlation length of the turbulence and $\eta_{b}$ ($\leq 1$)
defines the energy density ratio of the turbulent/mean field.
In the special case of $\alpha_{b}=1$, referred to as
the Bohm diffusion limit \citep{bohm49}, one gets the ordering
$\kappa_{{\parallel}_{b}}\sim\kappa_{\rm B}/\eta_{b}$,
where $\kappa_{\rm B}=cr_{\rm g}/3$ denotes the
Bohm diffusion coefficient for ultrarelativistic particles.
Considering the energy accumulation range of smaller $k_{{\parallel}_{b}}$
for the Kolmogorov turbulence with $\alpha_{b}=5/3$, \citet{zank98}
derived a modified coefficient that recovered the
scaling of equation~(\ref{eqn:27}) in the region of
$r_{\rm g}\ll k_{{\parallel}_{b},{\rm min}}^{-1}$.
As for the more complicated perpendicular diffusion, a phenomenological
hard-sphere scattering form of the coefficient is
$\kappa_{{\perp}_{b}}=\eta_{b}^{2}\kappa_{{\parallel}_{b}}$ in the
Bohm diffusion limit; and \citet{jokipii87} suggested a somewhat
extended version, $\kappa_{{\perp}_{b}}=\kappa_{{\parallel}_{b}}
/[1+(\lambda_{{\parallel}_{b}}/r_{\rm g})^{2}]$
(referred to as $\kappa_{\rm J}$ below), where
$\lambda_{{\parallel}_{b}}$ is the parallel mean free path (mfp).
A significantly improved theory of perpendicular diffusion
has recently been proposed by \citet{matthaeus03}, including
nonlinearity incorporated with the two-dimensional wavevector
$k_{{\perp}_{b}}$, whereupon for $\alpha_{b}=5/3$, \citet{zank04}
have derived an approximate expression of the corresponding
diffusion coefficient, although it still exhibits a somewhat
complicated form (referred to as $\kappa_{\rm Z}$).

On the other hand, within the present framework the gyroradius
of the injected energetic particles cannot be well defined,
because of $|\left<{\bf B}\right>|\simeq 0$ (\S\S~2.1 and 2.2).
Nonetheless, in order to make a fair comparison with the order
of the components of equation~(\ref{eqn:26}), the variable
${\bar B}$ is formally equated with $B=\left<|{\bf B}|^2\right>^{1/2}$.
In addition, the correlation length is chosen as
$k_{{\parallel}_{b},{\rm min}}\sim R^{-1}$,
corresponding to the setting in \S~2.3.
Then the ratio of $\kappa_{ii}$ for ${\tilde\kappa}=1$ to
equation~(\ref{eqn:27}) is found to take a value in the range of

\begin{equation}
{\kappa\over{\kappa_{{\parallel}_{b}}}}<\left(\alpha_{b}-1\right)
\left({1\over Z}{E\over {100~{\rm EeV}}}{{1~{\rm mG}}\over B}
{{100~{\rm pc}}\over{R}}\right)^{\alpha_{b}}
\label{eqn:28}
\end{equation}

\noindent
for the expected values of $\alpha$, $\beta\approx 2-3$.
Here $\kappa\equiv\kappa_{ii}$ and $q=Z|e|$ have been introduced.
Similarly, we get the scaling of
$\kappa/\kappa_{\rm B}\sim 10^{-1}(E/ZeBR)$, and
$\kappa/\kappa_{\rm J}\sim (E/ZeBR)^{1/3}$
for $\alpha_{b}=5/3$ and $\eta_{b}\sim 10^{-1}$
followed by $\lambda_{{\parallel}_{b}}\gg r_{\rm g}$.
Furthermore, considering the parameters given in
\citet{zank04}, which can be accommodated
with the above $\eta_{b}\sim 10^{-1}$, we also have
$\kappa/\kappa_{\rm Z}\sim (E/ZeBR)^{17/9}$ in the leading
order of $\kappa_{\rm Z}$, proportional to $r_{\rm g}^{1/9}$.
These scalings are valid for arbitrary species of
charged particles; for instance, setting $Z=1$ reflects
electrons, positrons, or protons (see footnote~1).
Particularly, for $\kappa <\kappa_{{\parallel}_{b}}$ in
equation~(\ref{eqn:28}), the efficiency of the present
DSA is expected to be higher than that of the conventional one
based on the simplistic QLT invoking parallel diffusion
\citep{biermann87}.
This can likely be accomplished for high-$Z$ particles,
as well as electrons with lower maximum energies.
Here note that $\kappa$ cannot take an unlimitedly smaller value
with decreasing $E$, since the effects of cold particle trapping in
the local magnetic fields make the approximation of no guide field
(eq.~[\ref{eqn:2}]) worse; and the lower limit of $E$ is relevant
to the injection condition called for the present DSA.
More on these is given in \S~3.

To apply the DSA model, one needs the effective
diffusion coefficient for the direction normal to the shock front,
referred to as the shock-normal direction.
For convenience, here we write down the coefficient for the
general case in which the current filaments are inclined by an
angle of $\phi$ with respect to the shock-normal direction.
In the tensor transformation of
$\kappa_{\mu\nu}^{\prime}=\Lambda_{\mu}^{\delta}\Lambda_{\nu}^{\epsilon}
\kappa_{\delta\epsilon}$, where

\begin{equation}
\kappa^{\prime}=\left(
\begin{array}{cc}
        \kappa_{11}^{\prime} & \kappa_{12}^{\prime} \\
        \kappa_{21}^{\prime} & \kappa_{22}^{\prime} \\
      \end{array}
\right),
\label{eqn:29}
\end{equation}
\begin{equation}
\Lambda=\left(
\begin{array}{cc}
        \cos\phi & -\sin\phi \\
        \sin\phi & \cos\phi \\
      \end{array}
\right),
\label{eqn:30}
\end{equation}

\noindent
we identify the shock-normal component $\kappa_{\rm n}$ with
$\kappa_{11}^{\prime}$.
It can be expressed as

\begin{equation}
\kappa_{{\rm n},\zeta}=\kappa_{\parallel ,\zeta}\left(
\cos^{2}\phi_{\zeta}+{\tilde\kappa}_{\zeta}\sin^{2}\phi_{\zeta}\right),
\label{eqn:31}
\end{equation}

\noindent
or simply as $\kappa_{{\rm n},\zeta}\approx\kappa_{\zeta}$ for
$\tilde\kappa_{\zeta}\approx 1$,
where the subscripts $\zeta={\rm I}$, ${\rm II}$ indicate
the upstream and downstream regions, respectively.
The expression of equation~(\ref{eqn:31}) appears to be the same
as equation~(4) in \citet{jokipii87}.
However, note again that now $\parallel$ and $\perp$ refer
to the direction of the linear current filaments,
compared to an astrophysical jet (\S~2.1 and Fig.~1).

\section[]{PARTICLE ACCELERATION BY SHOCK IN MAGNETIZED\\* CURRENT FILAMENTS}

We consider the particle injection mechanism
that makes the present DSA scenario feasible, retaining
the validity of the quasi-linear approximation.
Then, using the diffusion coefficient (eq.~[\ref{eqn:26}]), we estimate
the DSA timescale for arbitrary species of charged particles and
calculate, by taking the competitive energy loss processes into account, at
the achievable highest energies of the particles in astrophysical filaments.

\subsection[]{\it The Conception of Energy Hierarchy, Transition, and
Injection of\\* Cosmic-Ray Particles}

In the usual DSA context, equation~(\ref{eqn:31}) that
calls equation~(\ref{eqn:26}) determines the cycle time
for one back-and-forth of cosmic-ray particles across
the shock front, which is used below for evaluation of
the mean acceleration time \citep[\S~3.2;][]{gaisser90}.
Here we note that equation~(\ref{eqn:26}) is valid for
a high-energy regime in which the test particles with $E$ are unbound
to the local magnetic fields, so as to experience the nongyrating motion.
As shown below, this limitation can be deduced from the validity condition
of the quasi-linear approximation that has been employed in \S~2.2.
Using equations~(\ref{eqn:4}) and (\ref{eqn:7}), the validity condition
$\left< f_{\bf p}\right>\gg|\delta f_{\bf p}|$ can be rewritten as

\begin{equation}
\left< f_{\bf p}\right>\gg
\left|{q\over c}\int{\rm d}^2{\bf k}e^{i{\bf k}\cdot{\bf r}}
\left\{
\left({\bf k}\cdot{\bf v}\right)^{-1}
\left[{\bf v}\times\left({\bf k}\times{\bf A}_{\bf k}\right)\right]
\cdot{{\partial\left< f_{\bf p}\right>}\over{\partial{\bf p}}}
\right\}\right|,
\label{eqn:32}
\end{equation}

\noindent
where the off-resonance interaction with the quasi-static
fluctuations has been considered (\S~2.2).
For the momentum distribution function of
$\left<f_{\bf p}\right>\propto |{\bf p}|^{-\beta^{\prime}}$ for
the statistically accelerated particles with $E=|{\bf p}|c$
(\S~2.4), the RHS of equation~(\ref{eqn:32}) is of the order of
$\sim[|q{\bf A}({\bf r})|/(c|{\bf p}|)]\left<f_{\bf p}\right>$
for $\beta^{\prime}\sim O(1)$.
Therefore, we see that within the present framework, the
quasi-linear approximation is valid for the test particles
with an energy of $E\gg |qA({\bf r})|$, in a confinement region.
Note that this relation ensures the condition that
the gyroradius for the local field strength of $|{\bf B}({\bf r})|$
greatly exceeds the filament size (coherence length) of order
$\sim k^{-1}$, namely, $E/|q{\bf B}({\bf r})|\gg k^{-1}$
[equivalently, $E\gg |qA({\bf r})|$], except for a
marginal region of $k\sim R^{-1}$.
Obviously, this means that in the high-energy regime
of $E\gg |qA|$, the test particles are not strongly
deflected by a local magnetic field accompanying a fine
filament with its transverse scale of $\sim k^{-1}$.
On the other hand, in the cold regime of $E\ll |qA|$, the test
particles are tightly bound to a local magnetic field having the
(locally defined) mean strength, violating equation~(\ref{eqn:2}).
Here it is expected that the bound particles can diffuse along the
local field line, and hence, diffusion theories for a low-$\beta$ plasma
are likely to be more appropriate, rather than the present theory.

Summarizing the above discussions, there seem to exist
two distinct energy regimes for the test particles confined
in the system comprising numerous magnetized filaments: the
higher energy regime of $E\gg |qA|$, in which the particles
are free from the local magnetic traps, and the lower
energy regime of $E\ll |qA|$, in which the particles are bound
to the local fields, as compared to a low-$\beta$ state.
The hierarchy is illustrated in Figure~2, indicating
the characteristic trajectories of those particles.
When shock propagation is allowed, as seen in actual AGN jets,
the shock accelerator can energize the particles in each energy level.
At the moment, we are particularly concerned with EHE particle production
by a feasible scenario according to which energetic free particles,
unbound to small-scale
structure of the magnetized filaments, are further energized by the shock.
In this aspect, the particle escape from magnetically
bound states, due to another energization mechanism,
can be regarded as the injection of preaccelerated
particles into the concerned diffusive shock accelerator.
If the preaccelerator, as well, is of DSA, relying on the gyromotion
of bound particles \citep{drury83,biermann87,zank04}, the preaccelerator
also calls for the injection (in a conventional sense) in a far lower
energy level, owing to, e.g., the Maxwellian tail, or the energization
of particles up to the energies where the pre-DSA turns on
\citep[for a review, see][]{berezinskii90}.
The energy required for this injection, the so-called injection energy,
could be determined by, e.g., the competition with collisional resistance.
In order to distinguish from this commonly used definition of
``injection,'' we refer to the corresponding one, owing to the particle escape
from the magnetic traps, as the ``transition injection,'' in
analogy to the bound-free transition in atomic excitation.
The energy required to accomplish of the transition
injection is formally denoted as $E_{\rm inj}\sim |qA|_{\rm th}$,
where $|qA|_{\rm th}$ represents a threshold potential energy.
That is, the particles with $q$ and the energy exceeding
$E_{\rm inj}$ are considered to spaciously meander to
experience successive small deflection by the fields of
many filaments (Fig.~2), such that the present theory
is adequate for describing the particle diffusion.
This scattering property can be compared to that for the
conventional QLT in low-$\beta$ regimes: an unperturbed (zeroth order)
guiding center trajectory of gyrating particles bound to
a mean magnetic field must be a good approximation for
many coherence lengths of particle scatterer.

If both the injection and transition injection work,
the multistep DSA can be realized.
In the stage of $E\ll |qA({\bf r})|$, many acceleration scenarios
that have been proposed thus far (DSA: e.g., \citealt{drury83,biermann87};
shock drift acceleration: e.g., \citealt{webb83}; some versions of the
combined theories: e.g., \citealt{jokipii87,ostrowski88}; for a review,
see, e.g., \citealt{jones91}) can be candidates for the mechanism
of the preacceleration up to the energy range of
$E\sim |qA({\bf r})|$, although before achieving this energy level,
the acceleration, especially for electrons, might be,
in some cases, knocked down by the energy loss, such as
synchrotron cooling, collision with photons, and so on.
The relevant issues for individual specific situations
are somewhat beyond the scope of this paper
(observability is discussed in \S~4).
Here we just briefly mention that in the termination regions of large-scale
jets where the bulk kinetic energy is significantly converted into the
magnetic and particle energies, a conventional DSA mechanism involving
large-scale MHD turbulence might work up to EHE ranges
(Honda \& Honda [2004b] for an updated scenario of oblique
DSA of protons).

\subsection[]{\it Timescale of the Diffusive Shock Acceleration}

In the following, we focus on the DSA of energetic
free particles after the transition injection.
Let us consider a typical case of $\phi_{\rm I}=\phi_{\rm II}=0^{\circ}$
in equation~(\ref{eqn:31}), reflecting a reasonable situation that
a shock wave propagates along the jet comprising linear filaments.
Since the vectors of the random magnetic fields are on
the plane transverse to the current filaments, this
plane is perpendicular to the shock-normal direction.
That is, the shock across the perpendicular magnetic fields is considered.
In this case, no irregularity of magnetic surfaces in the
shock-normal direction exists, because of $k_{\parallel}=0$.
However, this does {\it not} mean that the particle flux diffusively
across the shock surface is in free-streaming;
note that the particles crossing the local fields with
nonsmall pitch angles suffer the orthogonal deflection.

Anyhow, the injected particles are off-resonantly scattered by
the filamentary turbulence, to diffuse, migrating back and forth
many times between the upstream and downstream regions of the shock.
As a consequence, a small fraction of them can be
stochastically accelerated to very high energy.
This scenario is feasible, as long as the filamentary
structure can exist around the discontinuity, as seen in
a kinetic simulation for shock propagation \citep{nishikawa03}
and an actual filamentary jet \citep{owen89}.
The timescale of this type of DSA is of the order of the
cycle time for one back-and-forth divided by the energy
gain per encounter with the shock \citep{gaisser90}.
Here the cycle time is related to the mean residence time of
particles (in regions I and II), which is determined by the
diffusive particle flux across the shock, dependent on
$\kappa_{{\rm n},\zeta}$ (eq.~[\ref{eqn:31}]).
For the moment, the shock speed is assumed to be nonrelativistic.
Actually, this approximation is reasonable, since the discrete knots
(for FR~I) and hot spots (for FR~II),
which are associated with shocks \citep[e.g.,][]{biretta83,carilli96},
preferentially move at a nonrelativistic speed, slower than the speed of
the relativistic jets \citep[e.g.,][]{meisenheimer89,biretta95}.
When taking the first-order Fermi mechanism into consideration
for calculation of the energy gain, the mean acceleration
time can be expressed as \citep{lagage83a,lagage83b,drury83}

\begin{equation}
t_{\rm acc}\simeq{3\over{U_{\rm I}-U_{\rm II}}}
\left({\kappa_{\rm n,I}\over U_{\rm I}}+
{\kappa_{\rm n,II}\over U_{\rm II}}\right),
\label{eqn:33}
\end{equation}

\noindent
where $U_{\rm I}$ and $U_{\rm II}$ are the flow speed of the
upstream and downstream regions in the shock rest frame, respectively.
The present case of $\phi_{\zeta}=0$ (in eq.~[\ref{eqn:31}])
provides $\kappa_{{\rm n},\zeta}=\kappa_{\parallel,\zeta}$, where
$\kappa_{\parallel,\zeta}$ is given in equation~(\ref{eqn:26}).
Here note the relation of $B_{\rm I}=B_{\rm II}$,
derived from the condition that the current density,
${\bf J}_{\zeta}\sim\nabla\times {\bf B}_{\zeta}$,
must be continuous across the shock front.
When assuming $\alpha_{\rm I}=\alpha_{\rm II}$ and
$\beta_{\rm I}=\beta_{\rm II}$, we arrive at the result

\begin{equation}
t_{a,{\rm acc}}\simeq{{3\sqrt{6}\pi\alpha r\left( r+1\right)}\over
{8\left(\alpha -1\right)\left(\beta_{a}+1\right)\left(\beta_{a}+2\right)
\left( r-1\right)}}{cE_{a}^{2}\over{q_{a}^{2}B^{2}RU^{2}}},
\label{eqn:34}
\end{equation}

\noindent
where the definitions of $\alpha\equiv\alpha_{\zeta}$,
$\beta_{a}\equiv\beta_{\zeta}$, $B\equiv B_{\zeta}$, and
$U\equiv U_{\rm I}=rU_{\rm II}$ have been introduced.
Equation~(\ref{eqn:34}) is valid for arbitrary species of particles ``$a$''
having energy $E_{a}$, spectral index $\beta_{a}$, and charge $q_{a}$.
Note that for the plausible ranges of the values of $\alpha$, $\beta_{a}$,
and $r$, the value of equation~(\ref{eqn:34}) does not significantly change.
The $\phi_{\zeta}$ dependence is also small, because of
${\tilde\kappa}\approx 1$, reflecting three-dimensional rms deflection of
unbound particles (\S~3.1 and Fig.~2).
In the scaling laws shown below, for convenience we use the
typical values of $\alpha=2$ \citep{montgomery79} and $r=4$
(for the strong shock limit), although we indicate, in equation~(\ref{eqn:39}),
the parameter dependence of the numerical factor.

\subsection[]{\it The Highest Energy of an Accelerated Ion}

In equation~(\ref{eqn:34}) for ions ($a=$''i''), we set $q_{\rm i}=Z|e|$
(see footnote~1) and $\beta_{\rm i}=3$ \citep[e.g.,][]{stecker99,demarco03}.
By balancing equation~(\ref{eqn:34}) with the timescale of the
most severe energy loss process, we derive the maximum possible
energy defined as $E_{\rm i,max}\equiv E_{\rm i}$.
In the environment of astrophysical filaments including extragalactic jets,
the phenomenological time balance equation can be expressed as

\begin{equation}
t_{{\rm i},{\rm acc}}={\rm min}\left(t_{\rm sh},t_{\rm i, syn},
t_{\rm n\gamma},t_{\rm nn^{\prime}}\right),
\label{eqn:35}
\end{equation}

\noindent
where $t_{\rm sh}$, $t_{\rm i,syn}$, $t_{\rm n\gamma}$,
and $t_{\rm nn^{\prime}}$ stand for the timescales of the shock propagation
(\S~3.3.1; eq.~[\ref{eqn:36}]), the synchrotron loss for ions
(\S~3.3.2; eq.~[\ref{eqn:40}]), the photodissociation of the nucleus
(\S~3.3.3; e.g., eq.~[\ref{eqn:43}]), and the collision of nucleus
``${\rm n}$'' with target nucleus ``${\rm n}^{\prime}$''
(\S~3.3.4; e.g., eq.~[\ref{eqn:46}]), respectively.
In addition, the energy constraint ascribed to the
spatial scale, i.e., the quench caused by the particle
escape, should also be taken into account (\S~3.3.5).
The individual cases are investigated below.

\subsubsection[]{\it The Case Limited by the Shock Propagation Time}

In the actual circumstances of astrophysical jets,
the propagation time of a shock through the jet, $t_{\rm sh}$,
restricts the maximum possible energy of accelerated particles.
The shock propagation time may be interpreted as the age of knots or
hot spots \citep{hh04b}, which can be crudely estimated as
$\sim L/U_{\rm prop}$, where $L$ represents a distance from
the central engine to the knot or hot spot being considered
and $U_{\rm prop}$ an average speed of their proper motion.
When assuming $U\sim U_{\rm prop}$, we get the scaling

\begin{equation}
t_{\rm sh}\sim 1\times 10^{11}{L\over{1~{\rm kpc}}}{{0.1c}\over{U}}~~{\rm s}.
\label{eqn:36}
\end{equation}

\noindent
For the case in which the shock is currently alive as is observed in AGN jets,
$t_{\rm sh}$ cannot be compared to the ``lifetime'' of the accelerator
that is considered in SNR shocks \citep[e.g.,][]{gaisser90}.

It is mentioned that in AGN jets, the timescale of
adiabatic expansion loss might be estimated as
$t_{\rm ad}\approx 3L/\left( 2\Gamma U_{r}\right)$,
where $\Gamma$ and $U_{r}$ represent the
Lorentz factor of jet bulk flows and the speed of
radial expansion, respectively \citep{muecke03}.
The fact that the jets are collimating well with an
opening angle of $\phi_{\rm oa}\lesssim 10\degr$ means
$U_{\rm prop}/U_{r}\gtrsim O(10)$; thereby,
$t_{\rm sh}\lesssim t_{\rm ad}$ for $\Gamma\lesssim{\cal O}(10)$.
Thus, it is sufficient to pay attention to the limit
due solely to the shock propagation time.
These circumstances are also in contrast with those in the SNRs,
where the flows are radially expanding without collimation,
and the shock propagation time (or lifetime) just reflects
the timescale of adiabatic expansion loss \citep[e.g.,][]{longair92}.

In equation~(\ref{eqn:35}), let us first consider
the case of $t_{\rm i, acc}=t_{\rm sh}$.
By equating (\ref{eqn:34}) with (\ref{eqn:36}), we obtain the following
expression for the maximum possible energy of an accelerated ion:

\begin{equation}
E_{{\rm i},{\rm max}}\sim 70~Z{B\over{1~{\rm mG}}}
\left({L\over{1~{\rm kpc}}}\right)^{1/2}
\left({R\over{100~{\rm pc}}}\right)^{1/2}
\left({U\over{0.1c}}\right)^{1/2}~~{\rm EeV}.
\label{eqn:37}
\end{equation}

\noindent
Note the ratio of $L/R\sim 360/(\pi\phi_{\rm oa}\sin\phi_{\rm va})\sim 10-100$
for the narrow opening angle of AGN jets of 
$\phi_{\rm oa}\sim 1\degr-10\degr$ and not-so-small viewing angle
$\phi_{\rm va}$ (e.g., for the M87 jet, $L\simeq 23R-33R$ for
$\phi_{\rm oa}\simeq 6\fdg 9$ [\citealt{reid89}] and
$\phi_{\rm va}=42\fdg 5\pm 4\fdg 5$ [\citealt{biretta95}]
or $30\degr-35\degr$ [\citealt{bicknell96}]).
Equation~(\ref{eqn:37}) (and eq.~[\ref{eqn:48}] shown below)
corresponds to the modified version of the simple scaling
originally proposed by \citet{hillas84}.

Concerning the abundance of high-$Z$ elements and their acceleration
to EHE regimes, the following points 1--4 may be worth noting:
\begin{enumerate}
\item Radial metallicity gradients are expected to be enhanced in
elliptical galaxies \citep[e.g.,][]{kobayashi04}.
Along with this, a significant increase of heavy elements has been
discovered in the central region of the nearby giant elliptical galaxy
M87 \citep{gastaldello02}, which contains a confirmed jet.
\item A variety of heavy ions including iron have been detected
in a microquasar jet \citep[SS\,433;][]{kotani96}.
\item The Haverah Park data favor proton primaries below
an energy of $\sim 50~{\rm EeV}$, whereas they appear to favor
a heavier composition above it \citep{ave00}.
\item The recent Fly's Eye data of $\sim 320~{\rm EeV}$
are compatible with the assumption of a hadron primary between
proton and iron nuclei \citep{risse04}.
\end{enumerate}
With reference to this observational evidence, we take the
possibility of acceleration of (or deceleration by) heavy particles
into consideration and indicate the charge ($Z$) and/or atomic number
($A$) dependence of the maximum possible energies and loss timescales.

\subsubsection[]{\it The Case Limited by the Synchrotron Cooling Loss}

The particles deflected by the random magnetic fields tend to emit
unpolarized synchrotron photons, which can be a dominant cooling process.
For relativistic ions, the timescale can be written as
$t_{\rm i, syn}\simeq 36\pi^{2}(A/Z)^{4}
[m_{\rm p}^{4}c^{7}/(e^{4}B^{2}E_{\rm i})]$,
where $m_{\rm p}$ denotes the proton rest mass \citep{gaisser90}.
In this expression, the energy of an accelerated ion, $E_{\rm i}$,
can be evaluated by equating $t_{\rm i,acc}$ with $t_{\rm i,syn}$.
That is, we have

\begin{equation}
{E_{\rm i}\over {Am_{\rm p}c^{2}}}=\xi(\alpha,\beta_{\rm i},r)
\left[{A\over Z^{2}}{{m_{\rm p}}\over{m_{\rm e}}}
{R\over r_{0}}\left({U\over c}\right)^{2}\right]^{1/3}.
\label{eqn:38}
\end{equation}

\noindent
Here the dimensionless factor $\xi$ is given by

\begin{equation}
\xi(\alpha,\beta,r)=\left[{{4\sqrt{6}\left(\alpha -1\right)
\left(\beta +1\right)\left(\beta +2\right)\left( r-1\right)}
\over{\alpha r\left( r+1\right)}}\right]^{1/3},
\label{eqn:39}
\end{equation}

\noindent
and $r_{0}=e^{2}/(4\pi m_{\rm e}c^{2})$ stands for the classical
radius of the electron, where $m_{\rm e}$ is the electron rest mass.
Substituting equation~(\ref{eqn:38}) into the expression of
$t_{\rm i, syn}$, the cooling timescale can be expressed as
a function of the physical parameters of the target object.
As a result, we find

\begin{equation}
t_{\rm i, syn}\sim 3\times 10^{15}
{1\over Z^{2/3}}\left({A\over{2Z}}\right)^{8/3}
\left({{1~{\rm mG}}\over B}\right)^{2}
\left({{100~{\rm pc}}\over{R}}\right)^{1/3}
\left({{0.1c}\over{U}}\right)^{2/3}~~{\rm s}.
\label{eqn:40}
\end{equation}

\noindent
Practically, this expression can be used in equation~(\ref{eqn:35})
for making a direct comparison with the other loss timescales.
For example, in the FR sources with $B\lesssim 1~{\rm mG}$
\citep{owen89,meisenheimer89,meisenheimer96,rachen93},
we have $t_{\rm i,syn}\gg t_{\rm sh}$,
so that the synchrotron cooling loss is ineffective.
It should, however, be noted that in blazars with $B\gtrsim 0.1~{\rm G}$
\citep{kataoka99,muecke01,aharonian02}, equation~(\ref{eqn:40}) becomes,
in some cases, comparable to equation~(\ref{eqn:36}).

When the equality of $t_{\rm i,acc}=t_{\rm i,syn}$ is fulfilled
in equation~(\ref{eqn:35}), equation~(\ref{eqn:38}) just provides
the maximum possible energy of the accelerated ion, which scales as

\begin{equation}
E_{\rm i,max}\sim 2A^{2/3}\left({A\over{2Z}}\right)^{2/3}
\left({R\over{100~{\rm pc}}}\right)^{1/3}
\left({U\over{0.1c}}\right)^{2/3}~~{\rm ZeV}.
\label{eqn:41}
\end{equation}

\noindent
The important point is that $t_{\rm i, acc}$ and $t_{\rm i, syn}$
are both proportional to $B^{-2}$, so that the
$B$ dependence of $E_{\rm i, max}$ is canceled out.
This property also appears in the case of electron acceleration
attenuated by the synchrotron cooling (\S~3.4.1).
In equation~(\ref{eqn:41}), it appears that for heavier ions,
$E_{\rm i, max}$ takes a larger value.
In the actual situation, however, the extremely energetic ions possess
a long mfp, and therefore, acceleration may be
quenched by the particle escape, as discussed in \S~3.3.5.

\subsubsection[]{\it The Case Limited by the Collision with Photons}

Here we focus on the proton-photon collision
that engenders a pion-producing cascade.
The characteristic time of the collision depends on the target
photon spectrum $n(\epsilon_{\rm ph})$ in the acceleration site, where
$n(\epsilon_{\rm ph})$ is the number density of photons per unit energy
interval for photon energy $\epsilon_{\rm ph}$.
For $n(\epsilon_{\rm ph})\propto \epsilon_{\rm ph}^{-2}$
\citep[e.g.,][]{bezler84}, typical for the FR sources \citep{rachen93},
the timescale can be expressed as
$t_{{\rm p}\gamma}\sim[u_{\rm m}/(\chi u_{\rm ph})]t_{\rm p,syn}$,
where $\chi\sim 200$ for the average cross section of
$\sigma_{\gamma {\rm p}}\sim 900~{\rm\mu barns}$ \citep{biermann87},
$u_{\rm ph}$ denotes the average energy density of target photons, and
$t_{\rm p,syn}=t_{\rm i,syn}|_{A=Z=1}$ (the subscript ``p'' indicates proton).
Thus, the expression of $t_{{\rm p}\gamma}$ includes
$E_{\rm p}$, i.e., the energy of the accelerated proton.
This can be evaluated by equating $t_{\rm p,acc}$
with $t_{{\rm p}\gamma}$, to have the form of

\begin{equation}
{E_{\rm p}\over {m_{\rm p}c^{2}}}={{\xi(\alpha,\beta_{\rm i},r)}\over{
\left(\chi\eta_{u}\right)^{1/3}}}
\left[{{m_{\rm p}}\over{m_{\rm e}}}{R\over r_{0}}
\left({U\over c}\right)^{2}\right]^{1/3},
\label{eqn:42}
\end{equation}

\noindent
where the definition $\eta_{u}\equiv u_{\rm ph}/u_{\rm m}$
has been introduced.
Substituting equation~(\ref{eqn:42}) into the expression of
$t_{{\rm p}\gamma}$, we obtain the following scaling of
the photomeson cooling time:

\begin{equation}
t_{{\rm p}\gamma}\sim 6\times 10^{15}
\left({200\over\chi}\right)^{2/3}\eta_{u}^{1/3}
{{10^{-10}~{\rm erg}~{\rm cm}^{-3}}
\over{u_{\rm ph}}}\left({{100~{\rm pc}}\over{R}}\right)^{1/3}
\left({{0.1c}\over{U}}\right)^{2/3}~~{\rm s}.
\label{eqn:43}
\end{equation}

\noindent
Note that for $\eta_{u}=\chi^{-1}\sim 5\times 10^{-3}$,
we have $t_{{\rm p}\gamma}=t_{\rm p, syn}$.

If the equality of $t_{\rm p,acc}=t_{{\rm p}\gamma}$ is
satisfied in equation~(\ref{eqn:35}), then equation~(\ref{eqn:42})
gives the maximum possible energy, which scales as

\begin{equation}
E_{\rm p, max}\sim 200
\left({200\over\chi}\right)^{1/3}
\left(1\over\eta_{u}\right)^{1/3}
\left({R\over{100~{\rm pc}}}\right)^{1/3}
\left({U\over{0.1c}}\right)^{2/3}~~{\rm EeV}.
\label{eqn:44}
\end{equation}

\noindent
For $\eta_{u}=\chi^{-1}$, equation~(\ref{eqn:44}) is
identical with equation~(\ref{eqn:41}) for $A=Z=1$.

\subsubsection[]{\it The Case Limited by the Collision with Particles}

The nucleus-nucleus collisions involving spallation reactions
can also be a competitive process in high-density regions.
For proton-proton collision, the timescale can be simply evaluated by
$t_{\rm pp^{\prime}}=(n_{\rm p^{\prime}}\sigma_{\rm pp^{\prime}}c)^{-1}$,
where $n_{\rm p^{\prime}}$ is the number density of target protons, and
$\sigma_{\rm pp^{\prime}}\approx 40~{\rm mbarns}$ denotes the cross section
in high-energy regimes.
The timescale can be rewritten as

\begin{equation}
t_{\rm pp^{\prime}}\simeq 8.3\times 10^{14}
{{1~{\rm cm}^{-3}}\over{n_{\rm p^{\prime}}}}~~{\rm s}.
\label{eqn:45}
\end{equation}

\noindent
It is found that for tenuous jets with $n_{\rm p^{\prime}}\ll 1~{\rm cm}^{-3}$,
the value of equation~(\ref{eqn:45}) is larger than the conceivable value
of equation~(\ref{eqn:36}); that is, the collisional loss is ineffective.

For the collision of an accelerated proton with a nonproton nucleus,
the timescale can be evaluated by the analogous notation,
$t_{\rm p{\rm N}^{\prime}}=(n_{A^{\prime}}\sigma_{{\rm p}A^{\prime}}c)^{-1}$,
where $n_{A^{\prime}}$ is the fractional number density of the
target nuclei having atomic number $A^{\prime}>1$.
Here we use an empirical scaling of the cross section,
$\sigma_{{\rm p}A^{\prime}}\approx\pi r_{0}^{2}A^{\prime 2/3}$,
where $r_{0}\simeq 1.4\times 10^{-13}~{\rm cm}$,
although the value of $r_{0}$ may be an overestimate for
very high energy collisions \citep[e.g.,][]{burbidge56}.
Combining $t_{\rm p{\rm N}^{\prime}}$ with $t_{\rm pp^{\prime}}$,
in general the timescale for collision of a proton with a nucleus
of an arbitrary composition can be expressed as

\begin{equation}
t_{\rm pn^{\prime}}\simeq 5.4\times 10^{14}
{1\over{0.65n_{\rm p^{\prime}}+
\sum_{A^{\prime}>1}n_{A^{\prime}}A^{\prime 2/3}}}~~{\rm s},
\label{eqn:46}
\end{equation}

\noindent
where $n_{\rm p^{\prime}}$ and $n_{A^{\prime}}$ are
both in units of ${\rm cm}^{-3}$.

In equation~(\ref{eqn:35}), we consider the case of
$t_{\rm p,acc}=t_{\rm pn^{\prime}}$.
By equating (\ref{eqn:34}) with (\ref{eqn:46}), we obtain the following
expression for the maximum possible energy of an accelerated proton:

\begin{equation}
E_{\rm p,max}\sim 200
\left({{100~{\rm cm}^{-3}}\over{n_{\rm p^{\prime}}+1.5
\sum_{A^{\prime}>1}n_{\rm A^{\prime}}A^{\prime 2/3}}}\right)^{1/2}
{B\over{1~{\rm mG}}}\left({R\over{100~{\rm pc}}}\right)^{1/2}
{U\over{0.1c}}~~{\rm EeV}.
\label{eqn:47}
\end{equation}

\noindent
As for the collision of an arbitrary accelerated nucleus with a target nucleus,
we can analogously estimate $t_{\rm nn^{\prime}}$ and $E_{\rm i,max}$.
In particular, the heavier nucleus--proton collision is more important,
since its timescale $t_{\rm np^{\prime}}$ is of the order of
$t_{\rm pp^{\prime}}/A^{2/3}$: for larger $A$ and $n_{\rm p^{\prime}}$,
it can be comparable to the other loss timescales.
For example, the parameters of $A=56$ (iron) and
$n_{\rm p^{\prime}}\sim 100~{\rm cm}^{-3}$ lead to
$t_{\rm np^{\prime}}\sim 4\times 10^{11}~{\rm s}$.
For the case of $t_{\rm i,acc}=t_{\rm np^{\prime}}$
in equation~(\ref{eqn:35}), we have the scaling of
$E_{\rm i,max}\sim 0.6Z^{2/3}(2Z/A)^{1/3}E_{\rm p,max}$,
where $E_{\rm p,max}$ is of equation~(\ref{eqn:47}) for
$n_{\rm p^{\prime}}\gg \sum_{A^{\prime}>1}n_{A^{\prime}}A^{\prime 2/3}$.

\subsubsection[]{\it Quenching by Particle Escape}

The particle escape also limits its acceleration; that is, the
spatioscale of the system brings on another energy constraint.
Relating to this point, in \S~2.4 we found the relation of
${\tilde\kappa}\approx 1$, meaning that the anisotropy
of the spatial diffusion coefficient is small.
It follows that the radial size of the jet (rather than $L$)
affects the particle confinement.
Recall here that in the interior of a jet the magnetic
field vectors tend to be canceled out, whereas around the envelope the
uncanceled, large-scale ordered field can appear \citep{hh04a}.
From the projected view of the jet, on both sides of the
envelope the magnetic polarities are reversed.

The spatially decaying properties of such an envelope field in the
external tenuous medium or vacuum might influence the transverse
diffusion of particles.
The key property that should be recalled is that
for $r\gg k^{-1}$ distant from a filament, the magnetic
field strength is likely to slowly decay, being proportional
to $\sim (kr)^{-1}$ \citep{honda00,honda02}.
It is, therefore, expected that as long as the radial size of the
largest filament, i.e., correlation length, is comparable to the
radius of the jet (\S~2.3), the long-range field pervades the
exterior of the jet, establishing the ``magnetotail'' with
the decay property of $\sim (k_{\rm min}r)^{-1}$ for
$r\gg k_{\rm min}^{-1}(\sim R)$.
In fact, in a nearby radio galaxy, the central kiloparsec-scale
``hole'' of the inner radio lobe containing a jet is filled with an
ordered, not-so-weak (rather strong) magnetic field of the order of
$10-100~{\rm \mu G}$ \citep{owen90}, whose magnitude is comparable
to (or $\sim 10~\%$ of) that in the jet \citep{owen89,heinz97}.
Presumably, the exuding magnetic field plays an additional role in
confining the leaky energetic particles with their long mfp of
$\lambda_{\perp}(\sim c\psi_{2}/\nu_{22})\sim R$.

In this aspect, let us express an effective confinement radius as
$R_{\rm c}={\tilde \rho}R$, where ${\tilde \rho}\gtrsim 1$,
and impose the condition that the accelerator operates for the particles
with the transverse mfp of $\lambda_{\perp}\leq R_{\rm c}$.
Then the equality gives the maximum possible energy in the form of

\begin{equation}
E_{\rm i,max}\sim 200~Z{\tilde\rho}^{1/2}
{B\over{1~{\rm mG}}}{R\over{100~{\rm pc}}}~~{\rm EeV}.
\label{eqn:48}
\end{equation}

\noindent
Values of $E_{\rm i,max}$ (and $E_{\rm p,max}$)
derived from the time balance equation~(\ref{eqn:35})
cannot exceed that of equation~(\ref{eqn:48}).
It appears that equation~(\ref{eqn:48}) can be compared to
the energy scaling derived from, in the simplest model,
the maximum gyroradius in a uniform magnetic field.

\subsection[]{\it The Highest Energy of an Accelerated Electron}

In a manner simiar to that explained in \S~3.3, we find the generic
scaling for the achievable highest energy of electrons.
In equation~(\ref{eqn:34}) for electrons ($a=$''e''),
we set $q_{\rm e}=-|e|$ and $\beta_{\rm e}=2$
\citep[e.g.,][]{meisenheimer89,rachen93,wilson02}.
By balancing equation~(\ref{eqn:34}) with the timescale of
a competitive energy loss process, we derive the maximum
possible energy defined as $E_{\rm e,max}\equiv E_{\rm e}$.
The time balance equation can be written as

\begin{equation}
t_{{\rm e},{\rm acc}}={\rm min}\left(t_{\rm e, syn},
t_{\rm ic}, t_{\rm br}\right),
\label{eqn:49}
\end{equation}

\noindent
where $t_{\rm e, syn}$, $t_{\rm ic}$, and $t_{\rm br}$ stand for
the timescales of the synchrotron loss for electrons
(\S~3.4.1; eq.~[\ref{eqn:51}]), the inverse Compton scattering
(\S~3.4.2; eq.~[\ref{eqn:54}]), and the bremsstrahlung emission loss
(\S~3.4.3; eq.~[\ref{eqn:57}]), respectively.
For positrons the method is so analogous that we omit the explanation.

\subsubsection[]{\it The Case Limited by the Synchrotron Cooling Loss}

For electrons, the synchrotron cooling is a familiar loss process,
and the timescale can be expressed as
$t_{\rm e,syn}\simeq 36\pi^{2}m_{\rm e}^{4}c^{7}/(e^{4}B^{2}E_{\rm e})$.
In this expression, the energy of an accelerated electron, $E_{\rm e}$,
can be evaluated by equating $t_{\rm e,acc}$ with $t_{\rm e,syn}$, to give

\begin{equation}
{E_{\rm e}\over{m_{\rm e}c^{2}}}=\xi(\alpha,\beta_{\rm e},r)
\left[{R\over r_{0}}\left({U\over c}\right)^{2}\right]^{1/3}.
\label{eqn:50}
\end{equation}

\noindent
Substituting equation~(\ref{eqn:50}) into the aforementioned expression
of $t_{\rm e,syn}$, the cooling timescale can be expressed
as a function of the physical parameters of the target object:

\begin{equation}
t_{\rm e,syn}\sim 1\times 10^{6}\left({{1~{\rm mG}}\over B}\right)^{2}
\left({{100~{\rm pc}}\over{R}}\right)^{1/3}
\left({{0.1c}\over{U}}\right)^{2/3}~~{\rm s}.
\label{eqn:51}
\end{equation}

\noindent
This can be used in equation~(\ref{eqn:49}) for
comparison with the other loss timescales.

When the equality of $t_{\rm e,acc}=t_{\rm e,syn}$ is
satisfied in equation~(\ref{eqn:49}), equation~(\ref{eqn:50})
gives the maximum possible energy, which scales as

\begin{equation}
E_{\rm e,max}\sim 50\left({R\over{100~{\rm pc}}}\right)^{1/3}
\left({U\over{0.1c}}\right)^{2/3}~~{\rm PeV}.
\label{eqn:52}
\end{equation}

\noindent
According to the explanation given in \S~3.3.2, equation~(\ref{eqn:52})
is independent of $B$ (see also eq.~[\ref{eqn:41}]).
The striking thing is that for plausible parameters,
the value of $E_{\rm e,max}$ is significantly larger
than that obtained in the context of the simplistic QLT
invoking the Alfv\'en waves \citep{biermann87}.
This enhancement is, as seen in equation~(\ref{eqn:28}), attributed
to the smaller value of the diffusion coefficient for electrons, which
leads to a shorter acceleration time, i.e., a smaller value of
equation~(\ref{eqn:33}), and thereby to a higher acceleration efficiency.

\subsubsection[]{\it The Case Limited by the Inverse Compton Scattering}

For the case of $\eta_{u}>1$, the inverse Compton scattering of accelerated
electrons with target photons can be a dominant loss process.
Actually, the environments of AGN jets often allow the synchrotron
self-Compton (SSC) and/or external Compton (EC) processes.
The characteristic time of the inverse Comptonization can be estimated as
$t_{\rm ic}\sim (t_{\rm e,syn}/\eta_{u})(\sigma_{\rm T}/\sigma_{\rm KN})$,
where $\sigma_{\rm T}=8\pi r_{0}^{2}/3$ and
$\sigma_{\rm KN}(\epsilon_{\rm ph}E_{\rm e})$
denote the total cross sections in the Thomson limit of
$\epsilon_{\rm ph}E_{\rm e}\ll m_{\rm e}^{2}c^{4}$ and the Klein-Nishina
regime of $\epsilon_{\rm ph}E_{\rm e}\gtrsim m_{\rm e}^{2}c^{4}$,
respectively \citep[e.g.,][]{longair92}.
The expression of $t_{\rm ic}$ includes $E_{\rm e}$, which is determined by
numerically solving the balance equation of $t_{\rm e,acc}=t_{\rm ic}$.
Because of $\sigma_{\rm KN}\leq\sigma_{\rm T}$,
the value of $E_{\rm e}$ is found to be in the region of

\begin{equation}
{E_{\rm e}\over {m_{\rm e}c^{2}}}\geq
{{\xi(\alpha,\beta_{\rm e},r)}\over{\eta_{u}^{1/3}}}
\left[{R\over r_{0}}\left({U\over c}\right)^{2}\right]^{1/3},
\label{eqn:53}
\end{equation}

\noindent
in the whole range of $\epsilon_{\rm ph}$.
Note that the equality in equation~(\ref{eqn:53}) reflects
the Thomson limit of $\sigma_{\rm KN}/\sigma_{\rm T}=1$.
Substituting the value of $E_{\rm e}$ into the expression of $t_{\rm ic}$,
we can evaluate the scattering time, which takes a value in the range of

\begin{equation}
t_{\rm ic}\geq 5\times 10^{8}\eta_{u}^{1/3}
{{10^{-10}~{\rm erg}~{\rm cm}^{-3}}
\over{u_{\rm ph}}}\left({{100~{\rm pc}}\over{R}}\right)^{1/3}
\left({{0.1c}\over{U}}\right)^{2/3}~~{\rm s}.
\label{eqn:54}
\end{equation}

\noindent
For a given parameter $\eta_{u}$, the larger value of
$u_{\rm ph}$ depresses the lower bound of $t_{\rm ic}$,
though the Klein-Nishina effects prolong the timescale.
It should be noted that the evaluation of $t_{\rm ic}$ along
equation~(\ref{eqn:54}) is, in equation~(\ref{eqn:49}),
meaningful only for $\eta_{u}\geq 1$; that is, the
relation of $\eta_{u}<1$ ensures $t_{\rm ic}>t_{\rm e,syn}$.

For the case of $t_{\rm e,acc}=t_{\rm ic}$ in equation~(\ref{eqn:49}),
$E_{\rm e}$, conforming to equation~(\ref{eqn:53}), gives the
maximum possible energy, which takes the value of

\begin{equation}
E_{\rm e,max}\geq 50\left({1\over{\eta_{u}}}\right)^{1/3}
\left({R\over{100~{\rm pc}}}\right)^{1/3}
\left({U\over{0.1c}}\right)^{2/3}~~{\rm PeV}
\label{eqn:55}
\end{equation}

\noindent
for $\eta_{u}\geq 1$.
Again, note that the Thomson limit sets the lower bound of $E_{\rm e,max}$.
It is found that the Klein-Nishina effects enhance the value of
$E_{\rm e,max}$ in the regime of
$\epsilon_{\rm ph}\gtrsim m_{\rm e}^{2}c^{4}/E_{\rm e,max}$.
Note here that $E_{\rm e,max}$ cannot unlimitedly increase in actual
circumstances but tends to be limited by the synchrotron cooling.
Combining equation~(\ref{eqn:52}) with equation~(\ref{eqn:55}), therefore,
we can express the allowed domain of the variables as follows:

\begin{equation}
1\geq{E_{\rm e,max}\over{50~{\rm PeV}}}
\left({{100~{\rm pc}}\over R}\right)^{1/3}
\left({{0.1c}\over U}\right)^{2/3}
\geq{1\over{\eta_{u}^{1/3}}}.
\label{eqn:56}
\end{equation}

\noindent
Note that the upper bound reflects the synchrotron limit.
In the critical case of $\eta_{u}=1$ reflecting the energy equipartition,
the generic equation~(\ref{eqn:56}) degenerates into equation~(\ref{eqn:52}).

\subsubsection[]{\it The Bremsstrahlung Loss}

The bremsstrahlung emission of electrons in the Coulomb field of
nuclei whose charge is incompletely screened also affects the acceleration.
The timescale can be evaluated by the notation
$t_{\rm br}=(n_{Z^{\prime}}\sigma_{{\rm rad,e}Z^{\prime}}c)^{-1}$,
where $n_{Z^{\prime}}$ is the fractional number density of the target
nuclei having charge number $Z^{\prime}$
and $\sigma_{{\rm rad,e}Z^{\prime}}$ describes the radiation
cross section \citep[e.g.,][]{heitler54}.
When the screening effects are small,
for interaction with a heavy composite we have

\begin{equation}
t_{\rm br}\simeq 1.4\times 10^{16}
\left\{\left[22+{\rm ln}
\left(E_{\rm e}/1~{\rm PeV}\right)\right]
\sum_{Z^{\prime}}n_{Z^{\prime}}Z^{\prime 2}
\right\}^{-1}~~{\rm s},
\label{eqn:57}
\end{equation}

\noindent
where $n_{Z^{\prime}}$ is in units of ${\rm cm}^{-3}$.

In the peculiar environments of high density, enhanced metallicity, and
lower magnetic and photon energy densities, equation~(\ref{eqn:57})
may be comparable with equation~(\ref{eqn:51}) or (\ref{eqn:54}).
In ordinary AGN jets, however, the corresponding
physical parameters seem to be marginal.
Note also that the bremsstrahlung timescale for
ion-ion interactions is larger, by the order of
$(A^{2}/Z^{4})(m_{\rm p}/m_{e})^{2}\sim 10^{7}/Z^{2}$,
than the value of equation~(\ref{eqn:57}), and found to
largely exceed the value of equation~(\ref{eqn:36}),
namely, the age of the accelerator.
That is why the ion bremsstrahlung has been
excluded in equation~(\ref{eqn:35}).

\section[]{DISCUSSION AND SUMMARY}

The feasibility of the present model could be verified by
the measurement of energetic photons emanating from a
source, typically, bright knots in nearby AGN jets.
In any case, the electrons with energy $E_{\rm e,max}$,
given in equation~(\ref{eqn:56}), emit the most energetic
synchrotron photons, whose frequency may be estimated as
$\nu^{\ast}\sim (E_{\rm e,max}/m_{\rm e}c^{2})^{2}(eB/m_{\rm e}c)$,
where the mean field strength ${\bar B}$ has been compared to
the rms strength $B$.
For $E_{\rm e,max}\sim 10~{\rm PeV}$ as an example,
the frequencies of $\nu^{\ast}\gtrsim 10^{22}~{\rm Hz}$
are found to be achieved for $B\gtrsim 10~{\rm\mu G}$.
In the gamma-ray bands, however, the energy flux of photons
from the synchrotron originator is predicted to be often overcome
by that produced by the inverse Comptonization of target photons.
In this case, as far as the condition of
$\epsilon_{\rm ph}E_{\rm e,max}\gg (m_{\rm e}c^{2})^{2}$
is satisfied, the boosted photon energy is given by
$\epsilon_{\rm ph}^{\prime}\sim E_{\rm e,max}$, independent of
the target photon energy $\epsilon_{\rm ph}$, thereby
irrespective of SSC or ECs.
This is in contrast to another case of
$\epsilon_{\rm ph}E_{\rm e,max}\ll (m_{\rm e}c^{2})^{2}$,
in which one has $\epsilon_{\rm ph}^{\prime}\sim\epsilon_{\rm ph}
(E_{\rm e,max}/m_{\rm e}c^{2})^{2}$ dependent on $\epsilon_{\rm ph}$.
Apparently, for the extremely high energy ranges of $E_{\rm e,max}$
achieved in the present scheme, the former condition is more likely
satisfied for a wide range of $\epsilon_{\rm ph}$.
Therefore, in the circumstances that the source is nearby such that
collision with the cosmic infrared background, involving
photon-photon pair creation, is insignificant,
$\epsilon_{\rm ph}^{\prime}$ ($\sim E_{\rm e,max}$) just gives the
theoretical maximum of gamma-ray energy, although the Klein-Nishina
effects also take part in lowering the flux level.
This means, in turn, that a comparison of the $\epsilon_{\rm ph}^{\prime}$
value (multiplied by an appropriate Doppler factor) with the
observed highest energy of the Compton emissions might constitute
a method to verify the present DSA for electrons.

The case for this method is certainly solidified when the operation
of the transition injection (\S~3.1) is confirmed.
Making use of the inherent property that the synchrotron photons emitted by
electrons having an energy above $|eA({\bf r})|$ reduce their polarization,
the energy hierarchy can be revealed by the polarization measurements,
particularly, with wide frequency ranges and high spatioresolution.
According to the reasoning that the critical frequency above which the
measured polarization decreases, $\nu_{\rm c}({\bf r})$, ought to be of
the order of
$\sim [|eA({\bf r})|/m_{\rm e}c^{2}]^{2}[|e{\bf B}({\bf r})|/m_{\rm e}c]$,
the related coherence length can be estimated as
$k_{\rm c}^{-1}\sim c\{\nu_{\rm c}({\bf r})
[m_{\rm e}c/|e{\bf B}({\bf r})|]^{3}\}^{1/2}$.
Note that when the locally defined gyroradius reaches this critical scale,
the bound electrons are released.
In actual circumstances, $\nu_{\rm c}$ and the polarization for a fixed
frequency band are, if anything, likely to increase near the jet surface,
where the large-scale coherency could appear (\S~3.3.5).
This may be responsible for the results of the polarization measurement
of a nearby filamentary jet, which indicate a similar transverse
dependence \citep{capetti97}.
In the sense of
$E_{\rm e,max}/E_{\rm inj}|_{q=-|e|}\ll E_{\rm i,max}/E_{\rm inj}|_{q=Z|e|}$,
the transition injection condition for electrons is more
restrictive than that for ions.
Thus, observational evidence of the present DSA scenario for
energetic electrons will, if it is obtained, strongly suggest
that the same scenario operates for ion acceleration,
providing its finite abundance.

To summarize, we have accomplished the modeling of the
diffusive shock accelerator accompanied by the quasi-static,
magnetized filamentary turbulence that could be self-organized
via the current filamentation instability.
The new theory of particle diffusion relies on the following
conditions analogous to those for the conventional QLT:
(1) the test particles must not be strongly deflected by a fine filament
but suffer the cumulative small deflection by many filaments, and
(2) the transverse filament size, i.e., the coherence length of
the scatterer, is limited by the system size transverse to the filaments;
whereas, more importantly, it is dependent on neither the gyration,
the resonant scattering, nor the explicit limit of the weak turbulence.
We have derived the diffusion coefficient from the
quasi-linear type equation and installed it in a DSA model
that involves particle injection associated with the
bound-free transition in the fluctuating vector potential.
By systematically taking the conceivable energy
restrictions into account, some generic scalings of
the maximum energy of particles have been presented.
The results indicate that the shock in kiloparsec-scale jets
could accelerate a proton and heavy nucleus to
$10-100~{\rm EeV}$ and ${\rm ZeV}$ ranges, respectively.
In particular, for high-$Z$ particles, and electrons as well,
the acceleration efficiency is significantly higher than that
derived from a simplistic QLT-based DSA, as is
deduced from equation~(\ref{eqn:28}).
Consequently, the powerful electron acceleration to ${\rm PeV}$
ranges becomes possible for the plausible parameters.

We expect that the present theory can be, mutatis
mutandis, applied for solving the problem of particle transport
and acceleration in GRBs \citep{nishikawa03,silva03}.
The topic is of a cross-disciplinary field closely relevant
to astrophysics, high-energy physics, and plasma physics involving
fusion science; particularly, the magnetoelectrodynamics of filamentary
turbulence is subject to the complexity of ``flowing plasma.''
In perspective, further theoretical details might be
resolved, in part, by a fully kinetic approach allowing
multiple dimensions, which goes far beyond the MHD context.\\

\appendix
\section{CALCULATION OF THE INTEGRAL COMPONENTS
$I_{12}$, $I_{21}$, AND $I_{22}$}

For instruction, we write down the derivation of equations~(\ref{eqn:24})
and (\ref{eqn:25}) for the collisionless scattering of injected test
particles by magnetized current filaments having the configuration
illustrated in Figure~1.
Making use of equation~(\ref{eqn:10}), $I_{ij}$ for $ij\neq 11$
(eq.~[\ref{eqn:12}]) can be explicitly written as

\begin{equation}
I_{12}=-i{q^{2}\over c^{2}}
\int{\rm d}^{2}{\bf k}{\rm d}\omega|A|_{{\bf k},\omega}^{2}
\left({\bf k}\cdot{\bf v}\right)
{\partial\over{\partial p_{\parallel}}}
\left\{\left[\omega-\left({\bf k}\cdot{\bf v}\right)\right]^{-1}
v_{\parallel}\left({\bf k}
\cdot{\partial\over{\partial{\bf p}_{\perp}}}\right)
\left<f_{\bf p}\right>\right\},
\label{eqn:a1}
\end{equation}
\begin{equation}
I_{21}=-i{q^{2}\over c^{2}}
\int{\rm d}^{2}{\bf k}{\rm d}\omega|A|_{{\bf k},\omega}^{2}
v_{\parallel}\left({\bf k}\cdot
{\partial\over{\partial{\bf p}_{\perp}}}\right)
\left\{ \left[\omega-\left({\bf k}\cdot{\bf v}\right)\right]^{-1}
\left({\bf k}\cdot{\bf v}\right)
{\partial\over{\partial p_{\parallel}}}\left<f_{\bf p}\right>\right\},
\label{eqn:a2}
\end{equation}
\begin{equation}
I_{22}=i{q^{2}\over c^{2}}
\int{\rm d}^{2}{\bf k}{\rm d}\omega|A|_{{\bf k},\omega}^{2}
v_{\parallel}\left({\bf k}\cdot
{\partial\over{\partial{\bf p}_{\perp}}}\right)
\left\{ \left[\omega-\left({\bf k}\cdot{\bf v}\right)\right]^{-1}
v_{\parallel}\left({\bf k}
\cdot{\partial\over{\partial{\bf p}_{\perp}}}\right)
\left<f_{\bf p}\right>\right\},
\label{eqn:a3}
\end{equation}

\noindent
where we have used a standard correlation function
(eq.~[\ref{eqn:13}]) reflecting random magnetic fluctuations
on the transverse plane to the linear current filaments (see Fig.~1).
Recalling the causality principle and noticing that the real
part does not contribute to the integration, we get

\begin{equation}
I_{12}=-{{\pi q^{2}}\over c^{2}}
\int{\rm d}^{2}{\bf k}{\rm d}{\omega}|A|_{{\bf k},\omega}^{2}
\left({\bf k}\cdot{\bf v}\right)
{\partial\over{\partial p_{\parallel}}}
\left\{\delta\left[\omega-\left({\bf k}\cdot{\bf v}\right)\right]
v_{\parallel}\left({\bf k}
\cdot{\partial\over{\partial{\bf p}_{\perp}}}\right)
\left<f_{\bf p}\right>\right\},
\label{eqn:a4}
\end{equation}
\begin{equation}
I_{21}=-{{\pi q^{2}}\over c^{2}}
\int{\rm d}^{2}{\bf k}{\rm d}{\omega}|A|_{{\bf k},\omega}^{2}
v_{\parallel}\left({\bf k}\cdot
{\partial\over{\partial{\bf p}_{\perp}}}\right)
\left\{\delta\left[\omega-\left({\bf k}\cdot{\bf v}\right)\right]
\left({\bf k}\cdot{\bf v}\right)
{\partial\over{\partial p_{\parallel}}}\left<f_{\bf p}\right>\right\},
\label{eqn:a5}
\end{equation}
\begin{equation}
I_{22}={{\pi q^{2}}\over c^{2}}
\int{\rm d}^{2}{\bf k}{\rm d}{\omega}|A|_{{\bf k},\omega}^{2}
v_{\parallel}\left({\bf k}\cdot
{\partial\over{\partial{\bf p}_{\perp}}}\right)
\left\{\delta\left[\omega-\left({\bf k}\cdot{\bf v}\right)\right]
v_{\parallel}\left({\bf k}
\cdot{\partial\over{\partial{\bf p}_{\perp}}}\right)
\left<f_{\bf p}\right>\right\}.
\label{eqn:a6}
\end{equation}

\noindent
Again, we use an ad hoc equation~(16), valid for a quasi-static mode
that retains the narrow spectral peak around $\omega\sim 0$.
Assuming that the magnetic turbulence is isotropic on the
transverse plane, the angular average of
equations~(\ref{eqn:a4})--(\ref{eqn:a6}) is carried out.
Taking account of the off-resonant scattering of particles
by the quasi-static random fields gives

\begin{equation}
I_{12}\sim -{{16\pi q^{2}}\over c^{2}}v_{\perp}
{\partial\over{\partial p_{\parallel}}}
\left({v_{\parallel}\over v_{\perp}}\right)
{\partial\over{\partial p_{\perp}}}
\left<f_{\bf p}\right>
\int_{k_{\rm min}}^{k_{\rm max}}{{{\rm d}k}\over k}I_{k},
\label{eqn:a7}
\end{equation}
\begin{equation}
I_{21}\sim -{{16\pi q^{2}}\over c^{2}}v_{\parallel}
{\partial^{2}\over{\partial p_{\perp}\partial p_{\parallel}}}
\left<f_{\bf p}\right>
\int_{k_{\rm min}}^{k_{\rm max}}{{{\rm d}k}\over k}I_{k},
\label{eqn:a8}
\end{equation}
\begin{equation}
I_{22}\sim {{16\pi q^{2}}\over c^{2}}v_{\parallel}
{\partial\over{\partial p_{\perp}}}
\left({v_{\parallel}\over v_{\perp}}\right)
{\partial\over{\partial p_{\perp}}}
\left<f_{\bf p}\right>
\int_{k_{\rm min}}^{k_{\rm max}}{{{\rm d}k}\over k}I_{k},
\label{eqn:a9}
\end{equation}

\noindent
where the ordering and the definition of $I_{k}$ are the same
as those denoted in \S~2.2.
For a given momentum distribution function of
$\left<f_{\bf p}\right>\propto |{\bf p}|^{-(\beta +2)}$
for the test particles with an ultrarelativistic energy of
$E=|{\bf p}|c=c(p_{\parallel}^{2}+p_{\perp}^{2})^{1/2}$,
the partial derivatives can be estimated as

\begin{eqnarray}
{\partial\over{\partial p_{\parallel}}}
\left({v_{\parallel}\over v_{\perp}}\right)
{\partial\over{\partial p_{\perp}}}\left<f_{\bf p}\right>
&\sim&\left(\beta +2\right)
\left[\left(\beta +3\right)\psi_{1}^{2}-\psi_{2}^{2}\right]
{c^{2}\over E^{2}}\left<f_{\bf p}\right> \nonumber \\
&=&{\partial^{2}\over{\partial p_{\parallel}^{2}}}\left<f_{\bf p}\right>,
\label{eqn:a10}
\end{eqnarray}
\begin{eqnarray}
{\partial^{2}\over{\partial p_{\perp}\partial p_{\parallel}}}
\left<f_{\bf p}\right>
&\sim&\left(\beta +2\right)\left(\beta +4\right)\psi_{1}\psi_{2}
{c^{2}\over E^{2}}\left<f_{\bf p}\right> \nonumber \\
&=&{\partial\over{\partial p_{\perp}}}
\left({v_{\parallel}\over v_{\perp}}\right)
{\partial\over{\partial p_{\perp}}}\left<f_{\bf p}\right>,
\label{eqn:a11}
\end{eqnarray}

\noindent
where $\psi_{1}\equiv p_{\parallel}/|{\bf p}|$ and
$\psi_{2}\equiv p_{\perp}/|{\bf p}|$.
Using equations~(\ref{eqn:a10}) and (\ref{eqn:a11}), and
equation~(\ref{eqn:22}) concerning the $k$-space integration
for $k_{\rm max}\rightarrow\infty$, one can arrange
equations~(\ref{eqn:a7})--(\ref{eqn:a9}) in the form of
$I_{ij}\sim \nu_{ij}\left<f_{\bf p}\right>$ to obtain
the expressions of equations~(\ref{eqn:24}) and (\ref{eqn:25}).\\

\clearpage

\begin{figure}
\plotone{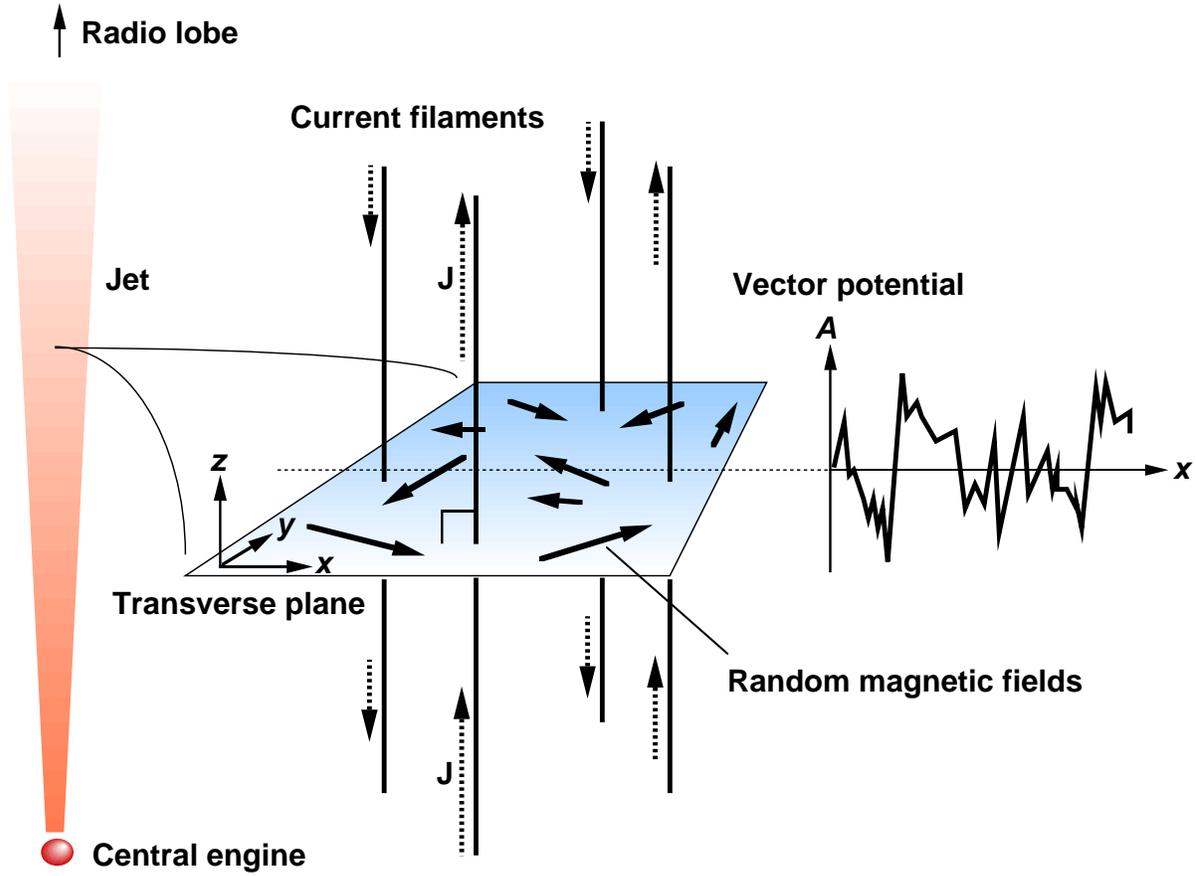}
\vspace{+1.0cm}
\caption{Schematics of the midscale configuration of
zeroth-order current density vectors ${\bf J}\sim J{\hat{\bf z}}$
({\it dotted arrows along the ``poles'' representing current filaments})
and magnetic field vectors ${\bf B}=(B_{x},B_{y},0)$ ({\it bold solid
arrows randomly distributed on the shaded plane}), and a spatial
profile of the vector potential ${\bf A}=A{\hat{\bf z}}$
({\it bold curve}), embodying the magnetized current filaments
that constitute the bulk of a large-scale jet.
Here the scalar $A(x,y)$ has been depicted as a function of
$x$ for a fixed $y$.
[{\it See the electronic edition of the Journal for a color
version of this figure.}]}
\end{figure}

\begin{figure}
\plotone{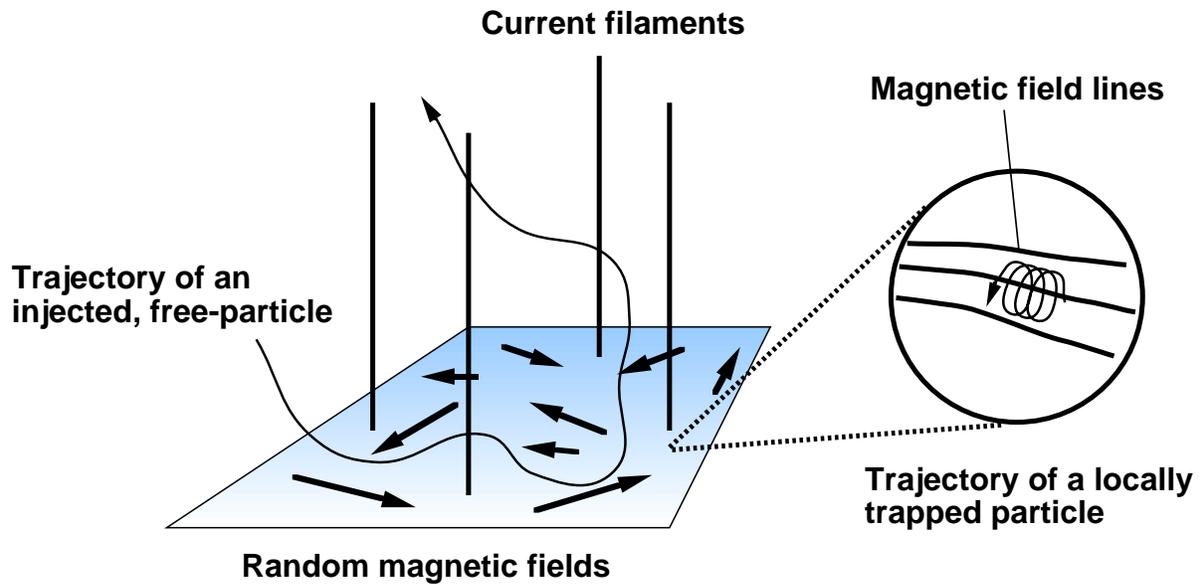}
\vspace{+0.5cm}
\caption{Schematics of the trajectories of a test free particle with
energy exceeding a threshold potential energy $|qA|_{\rm th}$,
and a gyrating, bound particle with energy below the local
potential $|qA(x,y)|$ ({\it light solid curves with arrow}).
In the energy hierarchy, the transition from the bound state to
the free state can be compared to the particle injection for the
present DSA, which requires the unbound particles to suffer
successive small deflections by the random magnetic fields.
For an explanation, see the text.
[{\it See the electronic edition of the Journal for a color
version of this figure.}]}
\end{figure}


\begin{thebibliography}{}
\bibitem[Abbasi et al.(2004a)]{abbasi04a}
Abbasi,~R.~U., et al. 2004a, \prl, 92, 151101

\bibitem[Abbasi et al.(2004b)]{abbasi04b}
---------. 2004b, \apj, 610, L73

\bibitem[Aharonian(2002)]{aharonian02}
Aharonian,~F.~A. 2002, \mnras, 332, 215

\bibitem[Appl \& Camenzind(1992)]{appl92}
Appl,~S., \& Camenzind,~M. 1992, \aap, 256, 354

\bibitem[Asada et al.(2000)]{asada00}
Asada,~K., Kameno,~S., Inoue,~M., Shen,~Z.-Q., Horiuchi,~S., \& Gabuzda,~D.~C.
2000, in Astrophysical Phenomena Revealed by Space VLBI,
ed. H.~Hirabayashi, P.~G.~Edwards, \& D.~W.~Murphy
(Sagamihara: ISAS), 51

\bibitem[Ave et al.(2000)]{ave00}
Ave,~M., Hinton,~J.~A., V\'azquez,~R.~A., Watson,~A.~A., \& Zas,~E. 2000,
\prl, 85, 2244

\bibitem[Bell(1978)]{bell78}
Bell,~A.~R. 1978, \mnras, 182, 147

\bibitem[Bell(2004)]{bell04}
---------. 2004, \mnras, 353, 550

\bibitem[Berezinski\u{i} et al.(1990)]{berezinskii90}
Berezinski\u{i},~V.~S., Bulanov,~S.~V., Dogiel,~V.~A., Ginzburg,~V.~L.,
Ptuskin,~V.~S. 1990, Astrophysics of Cosmic Rays (Amsterdam: North-Holland)

\bibitem[Bezler et al.(1984)]{bezler84}
Bezler,~M., Kendziorra,~E., Staubert,~R., Hasinger,~G., Pietsch,~W.,
Reppin,~C., Tr\"umper,~J., \& Voges,~W. 1984, \aap, 136, 351

\bibitem[Bicknell \& Begelman(1996)]{bicknell96}
Bicknell,~G.~V., \& Begelman,~M.~C. 1996, \apj, 467, 597

\bibitem[Biermann \& Strittmatter(1987)]{biermann87}
Biermann,~P.~L., \& Strittmatter,~P.~A. 1987, \apj, 322, 643

\bibitem[Biretta et al.(1983)]{biretta83}
Biretta,~J.~A., Owen,~F.~N., \& Hardee,~P.~E. 1983, \apj, 274, L27

\bibitem[Biretta et al.(1995)]{biretta95}
Biretta,~J.~A., Zhou,~F., \& Owen,~F.~N. 1995, \apj, 447, 582

\bibitem[Blandford(2000)]{blandford00}
Blandford,~R.~D. 2000, Phys.~Scr., T85, 191

\bibitem[Blandford \& Eichler(1987)]{blandford87}
Blandford,~R.~D., \& Eichler,~D. 1987, \physrep, 154, 1

\bibitem[Blandford \& Ostriker(1978)]{blandford78}
Blandford,~R.~D., \& Ostriker,~J.~P. 1978, \apj, 221, L29

\bibitem[Bohm(1949)]{bohm49}
Bohm,~D. 1949, in The Characteristics of Electrical Discharges in
Magnetic Fields, ed. A.~Guthrie \& R.~K.~Wakerling (New~York: McGraw-Hill), 77

\bibitem[Burbidge(1956)]{burbidge56}
Burbidge,~G.~R. 1956, \apj, 124, 416

\bibitem[Capetti et al.(1997)]{capetti97}
Capetti,~A., Macchetto,~F.~D., Sparks,~W.~B., \& Biretta,~J.~A.
1997, \aap, 317, 637

\bibitem[Carilli \& Barthel(1996)]{carilli96}
Carilli,~C.~L., \& Barthel,~P.~D. 1996, \aapr, 7, 1

\bibitem[Conway et al.(1993)]{conway93}
Conway,~R.~G., Garrington,~S.~T., Perley,~R.~A., \& Biretta,~J.~A.
1993, \aap, 267, 347

\bibitem[de~Marco et al.(2003)]{demarco03}
de~Marco,~D., Blasi,~P., \& Olinto,~A.~V. 2003, Astropart.~Phys., 20, 53

\bibitem[Drury(1983)]{drury83}
Drury,~L.~O'C. 1983, Rep.~Prog.~Phys., 46, 973

\bibitem[Farrar \& Biermann(1998)]{farrar98}
Farrar,~G.~R., \& Biermann,~P.~L. 1998, \prl, 81, 3579

\bibitem[Gabuzda(1999)]{gabuzda99}
Gabuzda,~D.~C. 1999, \nar, 43, 691

\bibitem[Gaisser(1990)]{gaisser90}
Gaisser,~T.~K. 1990, Cosmic Rays and Particle Physics
(Cambridge: Cambridge~Univ.~Press)

\bibitem[Gastaldello \& Molendi(2002)]{gastaldello02}
Gastaldello,~F., \& Molendi,~S. 2002, \apj, 572, 160

\bibitem[Greiner et al.(2003)]{greiner03}
Greiner,~J., et al. 2003, \nat, 426, 157

\bibitem[Greisen(1966)]{greisen66}
Greisen,~K. 1966, \prl, 16, 748

\bibitem[Heinz \& Begelman(1997)]{heinz97}
Heinz,~S., \& Begelman,~M.~C. 1997, \apj, 490, 653

\bibitem[Heitler(1954)]{heitler54}
Heitler,~W. 1954, The Quantum Theory of Radiation (London: Oxford~Univ.~Press)

\bibitem[Hillas(1984)]{hillas84}
Hillas,~A.~M. 1984, \araa, 22, 425

\bibitem[Honda(2000)]{honda00}
Honda,~M. 2000, Phys.~Plasmas, 7, 1606

\bibitem[Honda(2004)]{honda04}
---------. 2004, \pre, 69, 016401

\bibitem[Honda \& Honda(2002)]{honda02}
Honda,~M., \& Honda,~Y.~S. 2002, \apj, 569, L39

\bibitem[Honda \& Honda(2004a)]{hh04a}
---------. 2004a, \apj, 617, L37

\bibitem[Honda et al.(2000a)]{honda00a}
Honda,~M., Meyer-ter-Vehn,~J., \& Pukhov,~A. 2000a, Phys.~Plasmas, 7, 1302

\bibitem[Honda et al.(2000b)]{honda00b}
---------. 2000b, \prl, 85, 2128

\bibitem[Honda \& Honda(2004b)]{hh04b}
Honda,~Y.~S., \& Honda,~M. 2004b, \apj, 613, L25.

\bibitem[Jokipii(1987)]{jokipii87}
Jokipii,~J.~R. 1987, \apj, 313, 842

\bibitem[Jones \& Ellison(1991)]{jones91}
Jones,~F.~C., \& Ellison,~D.~C. 1991, \ssr, 58, 259

\bibitem[Junor et al.(1999)]{junor99}
Junor,~W., Biretta,~J.~A., \& Livio,~M. 1999, \nat, 401, 891

\bibitem[Kataoka et al.(1999)]{kataoka99}
Kataoka,~J., et al. 1999, \apj, 514, 138

\bibitem[Kazimura et al.(1998)]{kazimura98}
Kazimura,~Y., Sakai,~J.~I., Neubert,~T., \& Bulanov,~S.~V.
1998, \apj, 498, L183

\bibitem[Kobayashi(2004)]{kobayashi04}
Kobayashi,~C. 2004, \mnras, 347, 740

\bibitem[Kolmogorov(1941)]{kolmogorov41}
Kolmogorov,~A.~N. 1941, CR~Acad.~Sci.~URSS, 30, 301

\bibitem[Kotani et al.(1996)]{kotani96}
Kotani,~T., Kawai,~N., Matsuoka,~M., \& Brinkmann,~W. 1996, \pasj, 48, 619

\bibitem[Kraichnan(1965)]{kraichnan65}
Kraichnan,~R.~H. 1965, Phys.~Fluids, 8, 1385

\bibitem[Krymskii(1977)]{krymskii77}
Krymskii,~G.~F. 1977, Soviet~Phys.~Dokl., 22, 327

\bibitem[Lagage \& Cesarsky(1983a)]{lagage83a}
Lagage,~P.~O., \& Cesarsky,~C.~J. 1983a, \aap, 118, 223

\bibitem[Lagage \& Cesarsky(1983b)]{lagage83b}
---------. 1983b, \aap, 125, 249

\bibitem[Landau \& Lifshitz(1981)]{landau81}
Landau,~L.~D., \& Lifshitz,~E.~M. 1981, Physical Kinetics (Oxford: Pergamon)

\bibitem[Lee \& Lampe(1973)]{lee73}
Lee,~R., \& Lampe,~M. 1973, \prl, 31, 1390

\bibitem[Lobanov \& Zensus(2001)]{lobanov01}
Lobanov,~A.~P., \& Zensus,~J.~A. 2001, Science, 294, 128

\bibitem[Longair(1992)]{longair92}
Longair,~M.~S. 1992, High Energy Astrophysics, Vol.1: Particles, Photons
and Their Detection (Cambridge: Cambridge~Univ.~Press)

\bibitem[Lyutikov \& Blandford(2003)]{lyutikov03}
Lyutikov,~M., \& Blandford,~R.~D. 2003, preprint (astro-ph/0312347)

\bibitem[Matthaeus et al.(2003)]{matthaeus03}
Matthaeus,~W.~H., Qin,~G., Bieber,~J.~W., \& Zank,~G.~P.
2003, \apj, 590, L53

\bibitem[Medvedev \& Loeb(1999)]{medvedev99}
Medvedev,~M.~V., \& Loeb,~A. 1999, \apj, 526, 697

\bibitem[Meisenheimer et al.(1989)]{meisenheimer89}
Meisenheimer,~K., R\"oser,~H.-J., Hiltner,~P.~R., Yates,~M.~G., 
Longair,~M.~S., Chini,~R., \& Perley,~R.~A. 1989, \aap, 219, 63

\bibitem[Meisenheimer et al.(1996)]{meisenheimer96}
Meisenheimer,~K., R\"oser,~H.-J., \& Schl\"otelburg,~M. 1996, \aap, 307, 61

\bibitem[Montgomery \& Liu(1979)]{montgomery79}
Montgomery,~D., \& Liu,~C.~S. 1979, Phys.~Fluids, 22, 866

\bibitem[M\"ucke \& Protheroe(2001)]{muecke01}
M\"ucke,~A., \& Protheroe,~R.~J. 2001, Astropart.~Phys., 15, 121

\bibitem[M\"ucke et al.(2003)]{muecke03}
M\"ucke,~A., Protheroe,~R.~J., Engel,~R., Rachen,~J.~P., \&
Stanev,~T. 2003, Astropart.~Phys., 18, 593

\bibitem[Nishikawa et al.(2003)]{nishikawa03}
Nishikawa,~K.-I., Hardee,~P., Richardson,~G., Preece,~R., Sol,~H., \&
Fishman,~G.~J. 2003, \apj, 595, 555

\bibitem[Novak et al.(2003)]{novak03}
Novak,~G., et al. 2003, \apj, 583, L83

\bibitem[Olinto(2000)]{olinto00}
Olinto,~A.~V. 2000, \physrep, 333, 329

\bibitem[Ostrowski(1988)]{ostrowski88}
Ostrowski,~M. 1988, \mnras, 233, 257

\bibitem[Owen et al.(1990)]{owen90}
Owen,~F.~N., Eilek,~J.~A., \& Keel,~W.~C. 1990, \apj, 362, 449

\bibitem[Owen et al.(1989)]{owen89}
Owen,~F.~N., Hardee,~P.~E., \& Cornwell,~T.~J. 1989, \apj, 340, 698

\bibitem[Perley et al.(1984)]{perley84}
Perley,~R.~A., Dreher,~J.~W., \& Cowan,~J.~J. 1984, \apj, 285, L35

\bibitem[Potash \& Wardle(1980)]{potash80}
Potash,~R.~I., \& Wardle,~J.~F.~C. 1980, \apj, 239, 42

\bibitem[Rachen \& Biermann(1993)]{rachen93}
Rachen,~J.~P., \& Biermann,~P. 1993, \aap, 272, 161

\bibitem[Rawlings \& Saunders(1991)]{rawlings91}
Rawlings,~S., \& Saunders,~R. 1991, \nat, 349, 138

\bibitem[Reid et al.(1989)]{reid89}
Reid,~M.~J., Biretta,~J.~A., Junor,~W., Muxlow,~T.~W.~B., \& Spencer,~R.~E.
1989, \apj, 336, 112

\bibitem[Risse et al.(2004)]{risse04}
Risse,~M., Homola,~P., Gora,~D., Pekala,~J., Wilczynska,~B., \& Wilczynski,~H.
2004, Astropart.~Phys., 21, 479

\bibitem[Silva et al.(2003)]{silva03}
Silva,~L.~O., Fonseca,~R.~A., Tonge,~J.~W., Dawson,~J.~M., Mori,~W.~B., \&
Medvedev,~M.~V. 2003, \apj, 596, L121

\bibitem[Stecker \& Salamon(1999)]{stecker99}
Stecker,~F.~W., \& Salamon,~M.~H. 1999, \apj, 512, 521

\bibitem[Takeda et al.(1998)]{takeda98}
Takeda,~M., et al. 1998, \prl, 81, 1163

\bibitem[Tashiro \& Isobe(2004)]{tashiro04}
Tashiro,~M., \& Isobe,~N. 2004, Astron.~Herald, 97, 400

\bibitem[Teshima et al.(2003)]{teshima03}
Teshima,~M., et al. 2003, Proc. 28th Int. Cosmic-Ray Conf.
(Tsukuba), 437

\bibitem[Tsytovich \& ter~Haar(1995)]{tsytovich95}
Tsytovich,~V.~N., \& ter~Haar,~D. 1995,
Lectures on Non-linear Plasma Kinetics (Berlin: Springer)

\bibitem[Vall\'ee(2004)]{vallee04}
Vall\'ee,~J.~P. 2004, \nar, 48, 763

\bibitem[Waxman(1995)]{waxman95}
Waxman,~E. 1995, \prl, 75, 386

\bibitem[Webb et al.(1983)]{webb83}
Webb,~G.~M., Axford,~W.~I., \& Terasawa,~T. 1983, \apj, 270, 537

\bibitem[Wilson \& Yang(2002)]{wilson02}
Wilson,~A.~S., \& Yang,~Y. 2002, \apj, 568, 133

\bibitem[Yusef-Zadeh et al.(2004)]{yusefzadeh04}
Yusef-Zadeh,~F., Hewitt,~J., \& Cotton,~W. 2004, \apjs, 155, 421

\bibitem[Yusef-Zadeh \& Morris(1987)]{yusefzadeh87}
Yusef-Zadeh,~F., \& Morris,~M. 1987, \apj, 322, 721

\bibitem[Yusef-Zadeh et al.(1984)]{yusefzadeh84}
Yusef-Zadeh,~F., Morris,~M., \& Chance,~D. 1984, \nat, 310, 557

\bibitem[Zank et al.(2004)]{zank04}
Zank,~G.~P., Li,~G., Florinski,~V., Matthaeus,~W.~H.,
Webb,~G.~M., \& le~Roux,~J.~A. 2004, \jgr, 109, A04107

\bibitem[Zank et al.(1998)]{zank98}
Zank,~G.~P., Matthaeus,~W.~H., Bieber,~J.~W., \&
Moraal,~H. 1998, \jgr, 103, 2085

\bibitem[Zatsepin \& Kuzmin(1966)]{zatsepin66}
Zatsepin,~G.~T., \& Kuzmin,~V.~A. 1966, J.~Exp.~Theor.~Phys.~Lett., 4, 78\\

\end{thebibliography}
\end{document}